\documentclass[secnumarabic,superscriptaddress,amssymb, nobibnotes, aps, twocolumn, prl]{revtex4-2}
\usepackage{amsmath}
\usepackage{mathrsfs}  
\usepackage{mathtools} 
\usepackage{bbold}    
\usepackage{braket}    
\usepackage{gensymb}   

\usepackage{graphicx}      	
\graphicspath{{figures/}}
\usepackage{multirow}       
\usepackage{here} 			
\usepackage{url}		
\usepackage[plainpages=false,pdfpagelabels=true, linkcolor=cyan,breaklinks,colorlinks=true]{hyperref}  
\usepackage{soul}                   
\usepackage[dvipsnames]{xcolor}		
\usepackage{siunitx}
\usepackage{xspace}
\usepackage{booktabs}    
\usepackage{ulem}

\definecolor{atomictangerine}{rgb}{1.0, 0.6, 0.4}

\newcommand{\ute}{UTe\textsubscript{2}\xspace}

\newcommand{\Tc}{\ensuremath{T_{\rm c}}\xspace}
\newcommand{\Tco}{\ensuremath{T_{\rm c1}}\xspace}
\newcommand{\Tct}{\ensuremath{T_{\rm c2}}\xspace}
\newcommand{\Tcs}{\ensuremath{T_{\rm c}}'s\xspace}
\newcommand{\ctt}{\ensuremath{c_{33}}\xspace}
\newcommand{\att}{\ensuremath{\alpha_{33}}\xspace}
\newcommand{\Ps}{\ensuremath{P^{\star}}\xspace}
\newcommand{\kbT}{\ensuremath{k_B T}\xspace}
\newcommand{\kb}{\ensuremath{k_B}\xspace}

\begin{document}
	
	\title{Vanishing Phase Stiffness and Fluctuation-Dominated Superconductivity in \ute}%
	
	\author{Sahas Kamat}%
	
	\affiliation{Laboratory of Atomic and Solid State Physics, Cornell University, Ithaca, NY 14853, USA}
	\author{Jared Dans}
	\author{Shanta Saha}
	\affiliation{Maryland Quantum Materials Center, Department of Physics, University of Maryland, College Park, Maryland 20742, USA}
	\author{Daniel F. Agterberg}
	\affiliation{Department of Physics, University of Wisconsin–Milwaukee, Milwaukee, Wisconsin 53201, USA}
	\author{Johnpierre Paglione}
	\affiliation{Maryland Quantum Materials Center, Department of Physics, University of Maryland, College Park, Maryland 20742, USA}
	\affiliation{Canadian Institute for Advanced Research, Toronto, Ontario, Canada}
	\author{B. J. Ramshaw}
	\email{bradramshaw@cornell.edu}
	\affiliation{Laboratory of Atomic and Solid State Physics, Cornell University, Ithaca, NY 14853, USA}
	\affiliation{Canadian Institute for Advanced Research, Toronto, Ontario, Canada}

	\date{\today}%
	
	\begin{abstract}
		\bf
		
		Superconductivity in three dimensions is almost universally governed by Ginzburg-Landau mean field theory, with critical fluctuations typically confined to within a few percent of the transition temperature (\Tc). We report that the heavy-Fermion superconductor \ute exhibits superconducting fluctuations that extend over a temperature range as wide as \Tc itself---the largest observed for any three-dimensional superconductor. Through ultrasound measurements of the elastic moduli and sound attenuation, we find that \ute transitions from a mean-field-like state at ambient pressure to a fluctuation-dominated state at higher pressures. This regime is marked by elastic softening and an increase in sound attenuation that onsets well above \Tc, with the attenuation remaining anomalously high deep in the superconducting state. Our analysis suggests that these features stem from an extremely low superfluid phase stiffness. This results in a kinetic inductance as high as that of granular aluminum, but achieved in the clean limit. We propose a model where this exotic state is driven by dominant inter-band pairing mediated by ferromagnetic fluctuations, leading to `local' cooper pairs with a coherence length of only a few lattice constants.

	\end{abstract}
	
	\maketitle
	
	\section{Introduction}
	
	\ute is a heavy Fermion superconductor that hosts two distinct superconducting phases under hydrostatic pressure \cite{braithwaiteMultipleSuperconductingPhases2019}, as well as three (or more) superconducting phases in an applied magnetic field \cite{ranExtremeMagneticFieldboosted2019,knebelFieldReentrantSuperconductivityClose2019,ranExpansionHighFieldboosted2021,aokiFieldInducedSuperconductivitySuperconducting2021}. The high upper critical fields of these phases suggest that spin triplet pairing is likely involved, but the order parameter is still under debate \cite{theussSinglecomponentSuperconductivityUTe22024a, lewinReviewUTe2High2023}, leaving the pairing mechanism(s) still very much an open question. 
	
	Similarities in the field-temperature phase diagrams of \ute and the ferromagnetic superconductors UCoGe \cite{huySuperconductivityBorderWeak2007} and URhGe \cite{levyMagneticFieldInducedSuperconductivity2005} have led to speculation that ferromagnetic fluctuations might play an important role in pairing. While such fluctuations have been observed in high magnetic fields in \ute \cite{zambraGiantTransverseMagnetic2025}, the zero-field, ambient-pressure susceptibility is instead peaked at an antiferromagnetic wavevector \cite{knafoLowdimensionalAntiferromagneticFluctuations2021,duanIncommensurateSpinFluctuations2020}. Despite this, the fact that the high-pressure phase and the high-field SC2 phase are smoothly connected in the field-temperature-pressure phase diagram is highly suggestive that the pairing mechanism in these two phases is related \cite{vasinaConnectingHighFieldHighPressure2025}. In addition, first-principles calculations suggest that an intra-uranium-dimer ferromagnetic interaction is important for superconductivity for some region of parameter space \cite{xuQuasiTwoDimensionalFermiSurfaces2019}, which may be realized under high if not ambient pressures. 
	
	Whatever the (likely unconventional) pairing mechanism, the phenomenology of \ute is clearly well-described by Ginzburg–Landau mean field theory, at least at ambient pressure. For example, the specific heat anomaly at \Tc is sharp, with no signature of fluctuation effects. The applicability of mean field theory implies a strong superconducting phase stiffness, which is common among bulk metallic superconductors. A large phase stiffness may not, however, be universal to all superconducting phases of \ute. For example, the fact that the SC2 transition line changes slope dramatically at the SC1 phase boundary \cite{kamatThermodynamicDiscoveryTetracriticality2026}, while the slope of the SC1 phase boundary continues uninterrupted, suggests that SC2 is in some way less robust than SC1 (see \autoref{fig:ambientdata}). This is consistent with magnetic susceptibility measurements that suggest a larger London penetration depth (implying a smaller phase stiffness) in the SC2 phase \cite{wuMagneticSignaturesPressureInduced2025}. Any qualitative differences between the SC1 and SC2 phases that can be pinned down experimentally will be essential to sorting out the pairing mechanism, but to date there have been very few techniques applied under high pressures.

	Here, we perform ultrasonic measurements of the \ctt elastic modulus and the corresponding sound attenuation coefficient, $\alpha_{33}$, on \ute under pressure. We measure from ambient pressure up to $\approx$ 0.8 GPa and find that \ctt evolves from exhibiting a sharp thermodynamic discontinuity in the form of a ``jump'' in \ctt at \Tc---characteristic of a mean-field superconducting transition---to exhibiting softening of the elastic modulus over a broad temperature range above and below \Tc. This behaviour suggests that \ute evolves from a rather conventional mean-field superconductor at ambient pressure to a superconductor that is not well-described by Ginzburg-Landau (GL) mean field theory. The region over which we observe a breakdown of GL theory is larger than that of any known three-dimensional superconductor. In particular, we suggest that the SC2 phase is characterized by an anomalously low phase stiffness with `local' Cooper pairs whose coherence length is on the order of a few lattice constants. We introduce a simple model based on ferromagnetic fluctuations that leads to an inter-band pairing state with low phase stiffness.  
	
	\section{Experimental methods}
	
	\begin{figure*}[t]
		\centering
		\includegraphics[width=\textwidth]{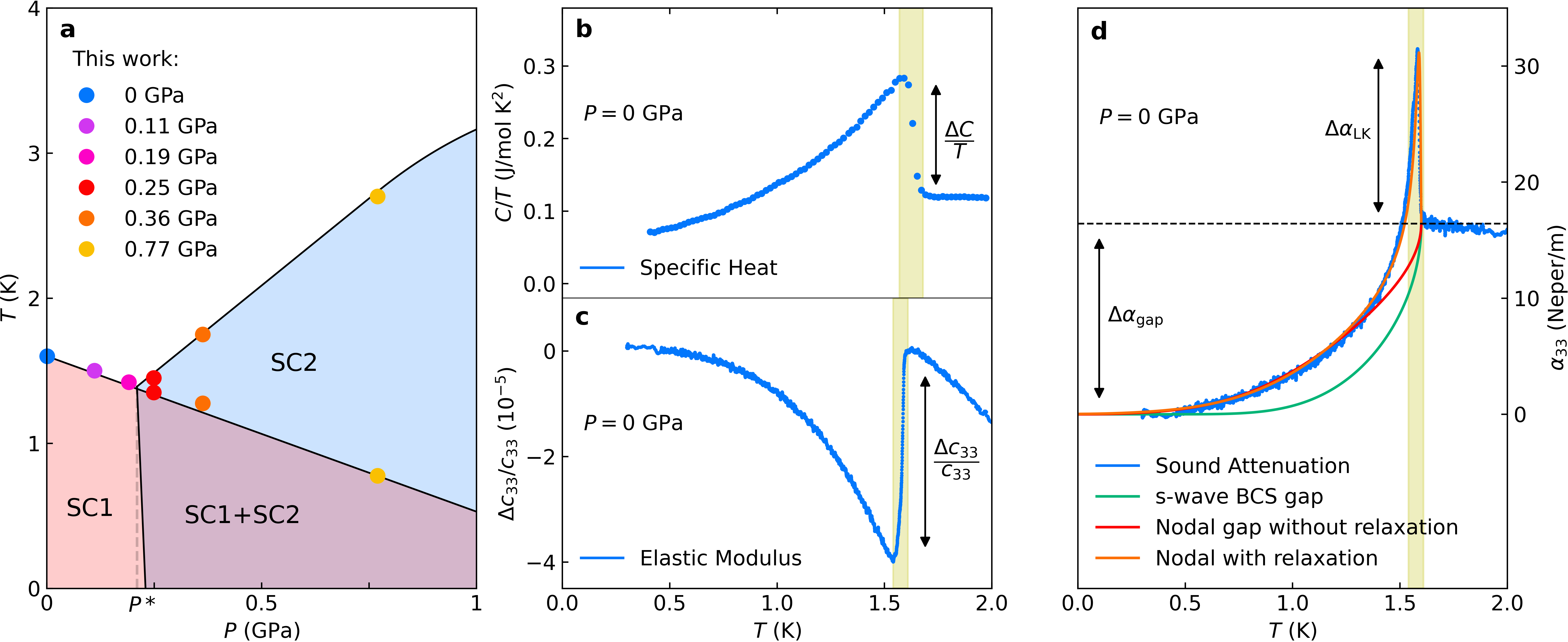}
		\caption{a) The temperature-pressure phase diagram of \ute in zero magnetic field. Phase boundaries (black lines) are guides to the eye adapted from Braithwaite et. al. and Kamat et. al. \cite{braithwaiteMultipleSuperconductingPhases2019,kamatThermodynamicDiscoveryTetracriticality2026}. The two superconducting phases, SC1 and SC2, are shaded  red and blue, respectively. Colored dots indicate transitions measured in the present work, and the pressure above which we observe two transitions is denoted by \Ps (\Ps is indicated by the dashed vertical line). b) Specific heat of \ute at ambient pressure as a function of temperature from \cite{ranNearlyFerromagneticSpintriplet2019}. A sharp upwards jump in $C/T$ is observed at \Tco upon entering the SC1 state. c) Ambient pressure $c_{33}$ elastic modulus as a function of temperature. A sharp downward jump $\Delta \ctt/\ctt$ is seen at \Tco. d) The longitudinal sound attenuation \att at ambient pressure as a function of temperature. An order-parameter relaxation peak $\Delta \alpha_\text{LK}$ is seen just below \Tc. As the quasiparticle density of states is gapped out below \Tc, we see the conventional decrease $\Delta \alpha_\text{gap}$ in the attenuation. Fits to different models of gap attenuation are shown.}
		\label{fig:ambientdata}
	\end{figure*}
	Single crystals of \ute were grown with the chemical vapour transport method described in \citet{ranNearlyFerromagneticSpintriplet2019} and \citet{ranExpansionHighFieldboosted2021}. Two samples, S1 and S2, were cut from a single bulk crystal for this study. The samples were aligned to better than 1\degree$~$ using X-ray diffraction performed in a Laue backscattering system. The crystals were then polished to produce two parallel faces normal to the (001) direction, and a 1 \si{\micro\meter} thick ZnO transducer was fabricated using the method described in \citet{theussSinglecomponentSuperconductivityUTe22024a}. The transducer excited ultrasonic waves with polarizations along the (100) and (001) directions, corresponding to shear ($c_{55}$) and longitudinal ($\ctt$) elastic moduli respectively. The longitudinal and shear signals were separated in the time domain using the known speeds of sound from \citet{theussAbsenceBulkThermodynamic2024a}. The relative changes in elastic moduli and the sound attenuation were determined using a traditional phase-comparison pulse echo technique (see \cite{theussSinglecomponentSuperconductivityUTe22024a}).
	
	Samples were pressurized in a piston-cylinder pressure cell with a 5:1 mixture of methanol and ethanol as a pressure medium. This medium was chosen because it is relatively hydrostatic over the range of pressures we measure\cite{klotzHydrostaticLimits112009}. Our ultrasound transducers probe a small ($100\ \si{\micro \meter}\times100\ \si{\micro \meter}\times300\ \si{\micro \meter}$) region of our sample, further decreasing the effects of any pressure gradients. The resistance of a manganin coil mounted inside the pressure cell was used to determine the pressure at room temperature. The pressure cell was heatsunk via a copper rod to the Helium-3 pot of an Oxford Instruments Heliox insert which was then cooled in a variable temperature insert in an Oxford Instruments Teslatron dry magnet system. This setup enabled measurements from 300 K down to 300 mK. At cryogenic temperatures, pressures were determined by comparing the change in \Tc to existing measurements \cite{braithwaiteMultipleSuperconductingPhases2019}. When more than one transition was measured, both $T_{c1}$ and $T_{c2}$ agreed well with existing data \cite{braithwaiteMultipleSuperconductingPhases2019,aokiMultipleSuperconductingPhases2020,vasinaConnectingHighFieldHighPressure2025}. We measured at six pressures in total. The transitions are shown as colored dots in \autoref{fig:ambientdata}a.

	\section{Results}
	
	The elastic modulus \ctt probes the sensitivity of the electronic free energy to compressional strain along the $c$ axis---$\epsilon_{zz}$---and was previously demonstrated to couple strongly to superconductivity in \ute \cite{theussSinglecomponentSuperconductivityUTe22024a}. Our \ute samples exhibits a single, sharp \Tc of 1.6 K at ambient pressure (the main text shows data on sample S1 except at $P=0.25$ GPa where we show data on S2; data on S2 at other pressures is shown in the S.I). \autoref{fig:ambientdata}b shows ambient-pressure ultrasound measurements on sample S1, and \autoref{fig:ambientdata}c shows specific heat measurements on a similarly grown sample. 
	
	Upon cooling through \Tc, the condensation of an order parameter leads to a sharp upwards jump $\Delta C/T$ in the specific heat. The \ctt elastic modulus has a corresponding downward jump $\Delta \ctt$ that is proportional to the heat capacity jump through an Ehrenfest relation:
	\begin{equation}
		\Delta \ctt = - \left( \partial T_c/ \partial \epsilon_{zz} \right)^2 \frac{\Delta C}{\Tc},
		\label{eq:mf33}
	\end{equation}
	where $\partial T_c/ \partial \epsilon_{zz}$ is the rate at which $T_c$ varies with $\epsilon_{zz}$ strain. The sharp jumps in both $C/T$ and \ctt at \Tc are expected for a mean-field-like phase transition.
	
	The sound attenuation coefficient \att probes how electronic quasiparticles relax when the Fermi surface is deformed by the strain $\epsilon_{zz}$ \cite{khanSoundAttenuationElectrons1987,thuillierMultipolarFermiSurface2025}. The most striking feature of the ambient-pressure data is the sharp peak in sound attenuation immediately below \Tc (\autoref{fig:ambientdata}d). While peaks in sound attenuation are not expected for conventional superconductors, similar peaks have been observed in UPt$_3$ \cite{bishopUltrasonicAttenuationUP$mathrmt_3$1984} and UBe$_{13}$ \cite{batlogg$ensuremathlambda$ShapedUltrasoundAttenuationPeak1985}. There, they were understood as coming from order-parameter relaxation due to a sign-changing superconducting gap \cite{miyakeLandauKhalatnikovDampingUltrasound1986}. A key characteristic of order parameter relaxation is that the peak height should scale linearly with ultrasound frequency \cite{miyakeLandauKhalatnikovDampingUltrasound1986}. We demonstrate this linear-in-frequency dependence in appendix II. This confirms that, while the peak is visually striking, it is only unconventional in the sense that it requires a sign-changing superconducting gap. Note that deep gap minima alone will not produce these peaks---the presence of these peaks upon entering both SC1 and SC2 constitute strong evidence that both states have sign-changing order parameters. Upon further cooling below \Tc, the superconducting gap opens and suppresses the quasiparticle density of states, producing a drop in sound attenuation as expected for a superconductor.
	
\begin{figure*}[t]
		\begin{center}
			\includegraphics[width=\textwidth]{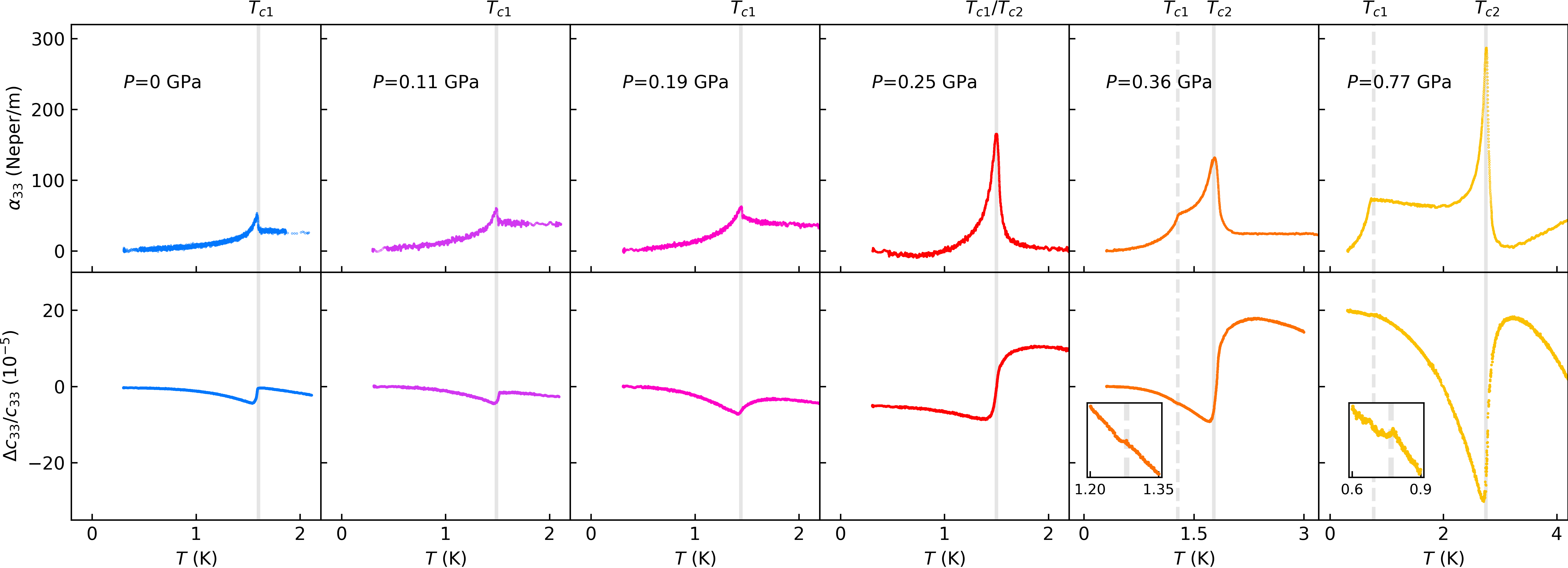}
		\end{center}
		\caption{Longitudinal sound attenuation \att and fractional change in elastic modulus $\Delta \ctt/\ctt$ as a function of temperature at different pressures. \Tco and \Tct are indicated with gray lines. \att exhibits a peak upon entering the SC1 state at the lowest three pressures ($P <\Ps$). At the higher three pressures ($P >\Ps$), the peak increases in magnitude and occurs upon entering the SC2 phase. The transition from the SC2 phase to the SC1 phase is clearly seen in the form of a kink as the attenuation decreases in the SC2 phase. Both features remain sharp at all pressures, indicating minimal pressure inhomogeniety (see \hyperref[supplement]{SI} for a quantitative analysis). 
			\ctt exhibits a sharp drop upon entering the SC1 state at the lowest two pressures, as expected for a superconducting transition. We see fluctuations onset at pressures greater than 0.19 GPa, in the form of a decrease in \ctt and increase in \att, at temperatures greater than \Tc. The size of the jump in \ctt increases as the pressure is increased above \Ps. The insets show a small jump that corresponds to the SC2-SC1 transition, which is seen at the highest two pressures. At the highest pressure of 0.77 GPa, the attenuation deep inside the SC2 superconducting state is larger than it is in the normal state, which is highly unusual for a superconductor.}
		\label{fig:pressure_dependence}
	\end{figure*}

	\autoref{fig:pressure_dependence} shows the evolution of the sound attenuation and elastic modulus as a function of temperature at six different pressures spanning ambient to 0.77 GPa (the pressure values are shown in \autoref{fig:ambientdata}a). At pressures greater than $\Ps = 0.2$ GPa, the peak in \att tracks the known phase boundary for SC2---we denote the transition to SC2 as \Tct, and the transition to SC1 as \Tco. 
	
	We first examine the evolution of sound attenuation with pressure. The peak in \att is roughly constant in amplitude for pressures less than the critical pressure $\Ps = 0.2$ GPa \cite{braithwaiteMultipleSuperconductingPhases2019}. The peak then increases substantially in magnitude for pressures larger than \Ps. At the highest two pressures, \att drops upon entering the SC1 state at \Tco. 
	
	There are several unexpected features in this data set. 1) For pressures near and above \Ps ($P\geq0.19$ GPa), \att increases as \Tc is approached in the normal state (i.e. from $T>\Tc$). 2) At the highest pressure, the attenuation in the superconducting state is substantially larger than in the normal state, even deep inside SC2, only dropping in a conventional manner at \Tco. Note that the peak in \att at \Tco remains sharp with increasing pressure for $P<\Ps$, as does the drop in \att at \Tco for $P>\Ps$. This indicates that there is minimal broadening of the transitions due to pressure inhomogeneity.
	
	Next, we examine the evolution of \ctt with pressure. The jump in \ctt is roughly constant in magnitude for pressures less than \Ps. Crossing to pressures higher than \Ps, the jump across the upper transition increases by a factor of three. This corresponds to an increase in $(d\Tc/d\epsilon_{zz})^2$ in \autoref{eq:mf33}, consistent with a more pressure-sensitive SC2 state with a larger slope of $d\Tc/dP$. At the two highest pressures, where both transitions are well separated in temperature, we find a small jump in \ctt at the lower transition (see insets to \autoref{fig:pressure_dependence}). 
	
	Concurrent with the increase in \att at temperatures above \Tc is the onset of softening in \ctt. Like the increase in \att above \Tc, this softening is not expected for a mean-field phase transition, and is suggestive of new physics associated with SC2. 

	\section{Analysis}
	\label{sec:analysis}
	
	
	\subsection{Quantifying the fluctuation region}
	
	Broad fluctuation regimes, while rare in superconductors, are not uncommon for other phase transitions. For example, magnetic systems typically show large fluctuation regions due to diverging spatial correlations approaching the phase transition \cite{theussStrongMagnetoelasticCoupling2022a}. The softening that we observe above \Tc in \ctt, and corresponding increase in \att, is therefore suggestive of spatial fluctuations in the magnitude and phase of the superconducting order parameter. To explore this idea, we extend the standard mean-field analysis to include small critical fluctuations of the order parameter. Here, we focus on modeling \ctt because it is a thermodynamic quantity and therefore easier to evaluate than \att, which is a transport quantity. 
	
\begin{figure}[h]
		\centering
		\includegraphics[width=\columnwidth]{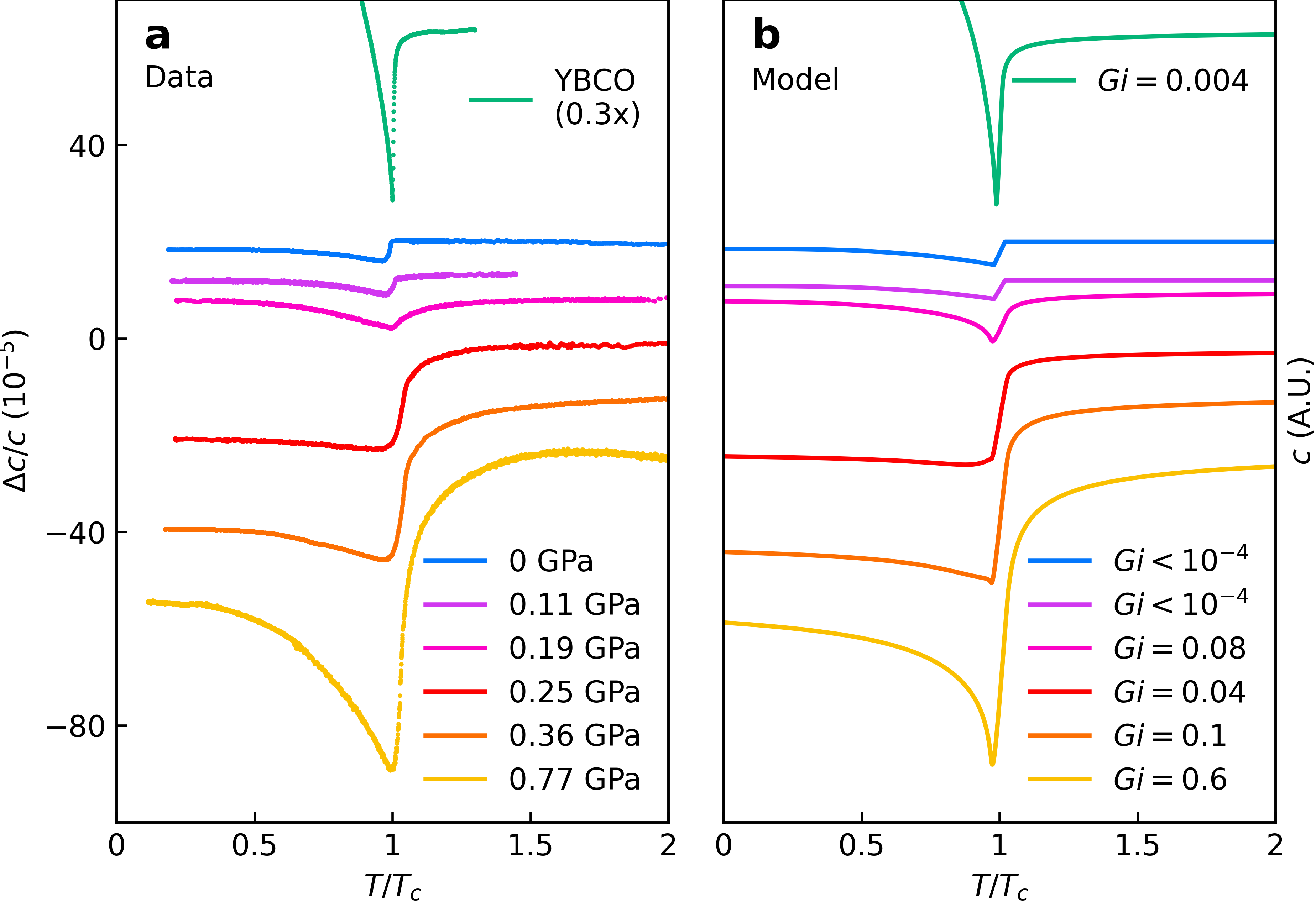}
		\caption{a) The elastic modulus \ctt across the superconducting transition as a function of temperature, with the background subtracted out. We also show the elastic response of YBa$_2$Cu$_3$O$_7$---a superconductor with a relatively large fluctuation region. Curves have been vertically offset for clarity. b) Theoretical predictions for \ctt using the Gaussian fluctuations model from the analysis section. The Ginzburg number $Gi$ quantifies the strength of the fluctuations and increases at higher pressures.}
		\label{fig:fits}
	\end{figure}

	First, to isolate the superconducting contribution to the temperature dependence of \ctt, we use the ambient pressure data for $T>\Tc$ as a background. We find that this background models the high temperature behavior of \ctt at all other pressures when scaled by a single pressure-dependent parameter. In the \hyperref[supplement]{SI}, we show that this parameter is approximately the Kondo coherence scale $T_K$, which manifests as a minimum in the temperature dependence of \ctt. We subtract the scaled background at each pressure and show the resulting curves in \autoref{fig:fits}a. For comparison, we also show data for YBa$_2$Cu$_3$O$_7$---a superconductor with a relatively large fluctuation region (of order a few percent of \Tc).
	
	Second, we analyze the contribution to \ctt with a model of Gaussian superconducting fluctuations. This model is intended to highlight the dependence of fluctuations on material parameters, such as phase stiffness, and is only a qualitative picture--- Gaussian fluctuations will fail to describe the critical behavior close to \Tc, where fluctuations are large. 
	
	The free energy of a superconductor, including contributions from a spatially-varying order parameter $\psi(\vec{r})$, is
	\begin{equation}
		F = \int\!\! d\vec{r}~\left[ f_n + \alpha\left(T\right) \left| \psi \right|^2 + \frac{1}{2}\beta \left| \psi \right|^4 + K \left| \nabla\psi \right|^2 - \gamma \epsilon_{zz}|\psi|^2\right],
		\label{eq:feng}
	\end{equation}
	where $f_n$ is normal-state free energy density including the elastic energy, $\alpha \equiv \alpha_0(T-\Tc)$ and $\beta$ are Landau coefficients, $K|\psi|^2 \equiv \kappa$ is the superfluid phase stiffness, and $\gamma \equiv \alpha_0 \frac{d \Tc}{d \epsilon_{zz}}$ is the strain-superconductivity coupling constant. \autoref{eq:feng} can be solved exactly in the limit of small fluctuations of the order parameter. This yields $\Delta\ctt$ in the vicinity of \Tc to be:
	\begin{equation}
		\Delta c_{33} = 
		-\left( \frac{d T_c}{d \epsilon_{zz}} \right)^2 \frac{\Delta C_{\text{MF}}}{T_C} \times 
		\begin{cases}
			\hspace{2.4em} \sqrt{\frac{Gi}{2t}} & T>\Tc \\
			\left(1 + \sqrt{\frac{Gi}{t}} \right) & T<\Tc.
		\end{cases}
		\label{eq:mfjump2}
	\end{equation}
	Here, $\Delta C_\text{MF}$ is the mean field heat capacity jump, $t = \left( T - T_c\right)/T_c$ is the reduced temperature, $k_B$ is the Boltzmann constant, and $Gi$ is the dimensionless Ginzburg number given by
	\begin{equation}
		Gi = \frac{ \Delta C_{\mathrm{MF}} \Tc}{\kappa_0^3} \left(\frac{k_B \Tc}{2\pi}\right)^2,
		\label{eq:ginzburg}
	\end{equation}
	where $\kappa_0$ is the phase stiffness at zero temperature. In the limit of large phase stiffness, $Gi$ vanishes and \autoref{eq:mfjump2} reduces to \autoref{eq:mf33}, i.e. there is no fluctuation contribution to \ctt when the phase stiffness is large. For a finite phase stiffness, a contribution that diverges as $1/\sqrt{t}$ turns on for both $T<T_c$ and $T>T_c$. 
	
	Mean field theory fails for temperatures close to \Tc given by $\big| T-\Tc\big| < \Tc \times Gi$, or $t<Gi$. In conventional superconductors, $Gi$ is of order $10^{-8}$ and the fluctuation region is never observed. Experimentally, we define the fluctuation region as the temperature range over which the fluctuation contribution to the heat capacity and \ctt is equal to its mean field value, i.e. when $t = Gi$. 
	
	

	\autoref{fig:fits}b plots \autoref{eq:mfjump2} with $Gi$ adjusted to best match the experimental data (see \hyperref[supplement]{SI} for details on how the fits are performed). For the two lowest pressures, we observe no precursor fluctuations and extract conventional Ginzburg numbers ($Gi < 10^{-4}$). Fluctuations onset at $P= 0.19$ GPa---slightly less than \Ps (this is confirmed by a single transition in $c_{55}$, see appendix I). The fluctuations grow as the pressure is increased until the $t^{-\alpha}$ contribution dominates the signal near \Tct at $P = 0.77$ GPa. 
	
	Note that the quantitative values of Ginzburg numbers extracted in this way should not be taken too literally---the central point of this analysis is to show that the data at higher pressures are modeled by an increasing fluctuation contribution that diverges at \Tc. Fits to a $t^{-1/2}$ power law obtained from Gaussian fluctuations are shown in the \hyperref[supplement]{SI}---the data fit qualitatively well, but the exact critical exponent may be different depending on the universality class of the transition. Critical fluctuations should also be observable in specific heat measurements, and recent quantitative specific heat measurements \cite{vasinaQuantitativeThermodynamicStudy2026} exhibit fluctuations consistent with those measured in this work. For a detailed comparison, see Appendix VI.
	
	
	\subsection{A model of superconductivity with low phase stiffness}
	
	\autoref{fig:fits}a shows a clear increase in the fluctuation contribution to \ctt at pressures greater than \Ps. This is evident in the raw data, which show both softening in \ctt and an enhancement of \att above \Tct. The fluctuation contribution is quantified by the large Ginzburg number we extract in \autoref{fig:fits}b.  
	
	From where do the fluctuations originate? The two lowest pressures show no fluctuation region associated with SC1. At $P=0.19$, fluctuations become apparent even though only SC1 is present---we attribute this to the close proximity of \Ps at this pressure (see appendix I). Above \Ps, increasing pressure increases the strength of fluctuations at \Tct, while there are no fluctuations at the lower \Tco. Clearly the fluctuations are associated with the SC2 phase.
	
	Why does the SC2 phase exhibit strong fluctuations of the order parameter while SC1 does not? At the level of the analysis above, this question becomes ``Why is $Gi$ large in SC2 and small in SC1''? \autoref{eq:ginzburg} suggests that a low superfluid phase stiffness in the SC2 phase could be responsible. 
	
	In a single-band BCS superconductor, the phase stiffness is simply the superfluid density divided by the quasiparticle effective mass. Naively, then, a low phase stiffness could be attributed to an anomalously small number of carriers, or to anomalously heavy carriers, condensing at \Tct. These explanations of our data face two difficulties: 1) the SC1 phase has an entirely conventional $Gi$, and there is no indication that the band structure of \ute changes dramatically with pressure; and 2) the anomalies in $C/T$ at \Tco and \Tct are within a factor of 2 of each other in size for pressures between $P = 0.38$ and $0.79$ GPa \cite{vasinaConnectingHighFieldHighPressure2025}, and therefore it cannot be that the mass (i.e. density of states) of the carriers condensing at \Tct is significantly different from those condensing at \Tco.
	
	
	Instead, we suggest that \textit{inter}-band pairing can produce an anomalously low phase stiffness. To demonstrate this, we consider pairing mediated by intra-dimer ferromagnetic fluctuations. Note that these fluctuations are distinct from the thermally-driven superconducting fluctuations we see in our data. Such fluctuations have been suggested to be relevant by first principle calculations \cite{xuQuasiTwoDimensionalFermiSurfaces2019} and are consistent with the dimer ferromagnetic configuration observed in spin fluctuations at ambient pressure \cite{Knafomagneticfluctuations2021} and in the magnetically ordered state at high pressure \cite{Knafomagneticorder2025}. These fluctuations are known to support a pairing state that is spin-triplet and site-singlet \cite{shishidouTopologicalBandSuperconductivity2021} (i.e. the Cooper pair orbital wavefunction is $\Psi(r_1,r_2) = 1/\sqrt{2}\left[\phi_1(r_1)\phi_2(r_2)-\phi_2(r_1)\phi_1(r_2)\right]$, where $\phi_1$ and $\phi_2$ are the states at the two uranium sites, and $r_1$ and $r_2$ are the electronic coordinates). When projected onto bands, this pairing state has both intra- and inter-band components because the two uranium states hybridize to form bonding and anti-bonding bands in the momentum basis. In most theoretical analyses of this pairing state \cite{shishidouTopologicalBandSuperconductivity2021,christiansen_nodal_2025,teiPairingSymmetries2024} the intra-band components dominate the pairing state. However, such a state would be expected to exhibit a usual BCS-like superfluid stiffness.
	
	Recently, Samokhin showed \cite{samokhinGinzburgLandauEnergyMultiband2024} that a dominant inter-band pairing state can emerge as the superconducting state with the highest $T_c$ if the following two conditions are satisfied: i) the interaction for the inter-band pairing component is the largest;  and ii) the inter-band component is Josephson coupled to the intra-band components.  Conditions i) and ii) are naturally satisfied from inter-dimer ferromagnetic fluctuations, where the relative weight of intra- to inter-band pairing is given by the ratio of intra- to inter-dimer hopping, $\epsilon$.  When $\epsilon<<1$, the intra-band pairing interaction is a factor of $\epsilon^2$ smaller than the inter-band pairing interaction. Taking the intra- and inter-dimer hopping parameters from \cite{theussSinglecomponentSuperconductivityUTe22024a}, where they were fitted to reproduce the the Fermi surface measured by quantum oscillations, we estimate $\epsilon^2 \approx 6\times10^{-3}$. This suggests that the inter-band pairing state could be dominant and produce a stable pairing state under favorable circumstances. 
	
\begin{figure}[h]
	\centering
	\includegraphics[width=.5\textwidth]{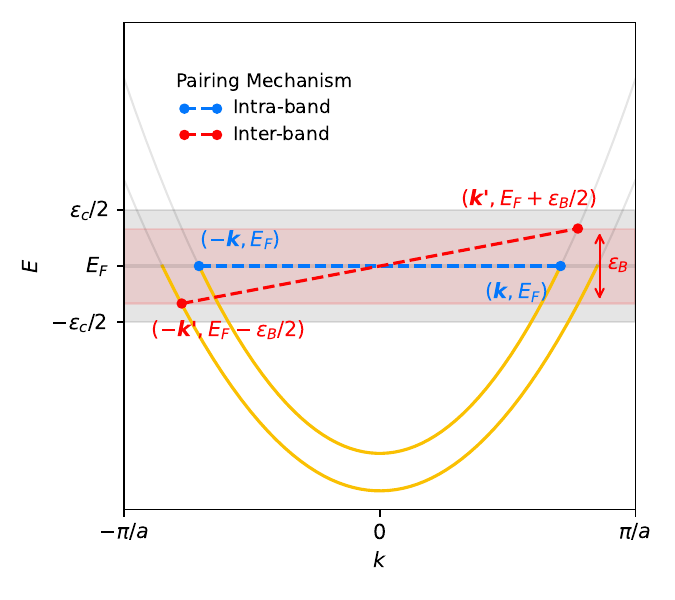}
	\caption{A cartoon depicting intra- and inter-- band pairing for a simple 2-band parabolic dispersion. Blue dots represent intra-band pairs at momenta $k$ and $-k$, and red dots represent inter-band pairs at momenta $k'$ and $-k'$. The intra-band pairs sit at the Fermi energy $E_F$. The gray shaded area represents the pairing energy shell. For an inter-band pairing state, this pairing energy ($\varepsilon_c$) must be greater than the band splitting $\varepsilon_B$ \cite{samokhinGinzburgLandauEnergyMultiband2024}, which is denoted by the red shaded area.}
	\label{fig:schematics}
\end{figure}

	This dominant inter-band pairing state has a superfluid stiffness that is much smaller than the more usual intra-band pairing states. This is because, for uniform superconducting states where $\vec{k}$ and $-\vec{k}$ are paired, conventional intraband pairs are formed at the Fermi energy, whereas interband pairs are necessarily formed at a finite energy difference (\autoref{fig:schematics}). Generically, then, the phase stiffness of the interband state is small compared to the phase stiffness of the intraband state by the larger of two possible factors. The first factor is $(k_{\rm B}\Tc/\mathcal{E}_b)^2$, where $\mathcal{E}_b$ is the energy splitting between the two bands, which is the intrinsic phase stiffness of the interband pairing state. The second factor is $\epsilon^2$, which emerges from the weak Josephson-coupled intra-band pairing contributions (see the appendix III for more details). In our case, $T_c$ is smaller than the inter-dimer hopping, so that the factor $\epsilon^2$ determines the phase stiffness. Again taking $\epsilon^2 \approx 6\times10^{-3}$ from \citet{theussSinglecomponentSuperconductivityUTe22024a}, we obtain a Ginzburg number for the inter-band state of $Gi_{\rm inter} \approx (5\times 10^{6})~ Gi_{\rm intra}$. This is consistent with the (minimum) four-order-of-magnitude increase in $Gi$ we observe when transitioning from SC1 to SC2 (independent estimates of $Gi$ at ambient pressure give $Gi \approx 10^{-5}$, see Appendix IV). It is worth noting that the increase in $Gi$ we observe with increasing pressure across the SC2 phase is consistent with the intra- and inter-dimer hopping ratios, $\epsilon$, changing by a factor of 2.4, which is not unreasonable. 

\section{Discussion}

Given the above analysis, we suggest that the primary distinguishing feature between SC1 and SC2 is that SC1 is an intra-band-paired superconductor, with a conventional, BCS-like phase stiffness, whereas SC2 is an inter-band-paired superconductor, with an anomalously low phase stiffness. This is consistent with the suggestion of an anomalously long penetration depth in the SC2 phase under pressure \cite{wuMagneticSignaturesPressureInduced2025}, and other works \cite{vijayanambikaEnhancementAntiferromagneticSpin2026,ambikaPossibleCoexistenceAntiferromagnetic2022,rosuelFieldInducedTuningPairing2023,weinbergerMetamagnetismUTe2Roles2026} that suggest different pairing mechanisms in the SC1 and SC2 phases. Note that this does \textit{not} imply a low density of Cooper pairs. Instead, the BCS/GL expression relating phase stiffness and superfluid density is no longer applicable in the inter-band state.


A large fluctuation region has direct implications for the spatial extent of the Cooper pairs, independent of the pairing mechanism. Intuition for this can be gained by examining magnetic phase transitions, where large Ginzburg numbers are not uncommon: $Gi\approx0.14$ in YbRh$_2$Si$_2$, for example \cite{krellner_violation_2009}. This departure from mean-field behaviour is associated with the short length scale of the magnetic exchange interaction (typically nearest or next-nearest neighbor), allowing for large spatial fluctuations of the order parameter. In superconductors, the relevant length scale is the coherence length, which is typically thousands of lattice sites. Using the GL definitions of phase stiffness and coherence length, a Ginzburg number of 0.6 in \ute at 0.77 GPa can be translated to a coherence length of only a few lattice constants: roughly 8 $\si{\angstrom}$. This ``local" Cooper pairing is in stark contrast with nearly all other superconductors, with the possible exception of underdoped cuprates \cite{blatterVorticesHightemperatureSuperconductors1994}. A short coherence length is consistent with the unusually large upper critical field ($H_\text{c2}>35$ T) of the SC2 state \cite{ranExtremeMagneticFieldboosted2019}.
We also note that an extended fluctuation regime can enable the existence of exotic precursor non-superconducting phases, such as the putative fermion quadrupling phase observed in Ba$_{1-x}$K$_x$Fe$_2$As$_2$ \cite{grinenkoStateSpontaneouslyBroken2021} To date we do not have any thermodynamic evidence for such a phase above $T_{c2}$.

To the best of our knowledge, the fluctuation regime shown in \autoref{fig:fits} at $P = 0.77$ GPa is larger than in any other three-dimensional superconductor, emphasizing the truly unique nature of the SC2 phase in \ute. The behavior of the sound attenuation at this pressure is particularly striking---even below the sharp peak at \Tct, the attenuation remains significantly higher in the superconducting state as compared to the normal state. It will be interesting to probe this region of the phase diagram with other dynamic probes, such as microwave conductivity and NMR. As we demonstrate in Appendix V, the low phase stiffness of SC2 at $P = 0.77$ GPa translates to a kinetic inductance rivaling that of granular aluminum and NbTiN \cite{bretz-sullivanHighKineticInductance2022}. Stabilizing the SC2 phase using epitaxial strain could provide an interesting route to building high kinetic inductance superconducting elements in the clean limit.


	
	\newpage
	
	\section{Acknowledgments}
	
	Research at Cornell was supported by the Department of Energy, Office of Basic Energy Sciences Award No. DE-SC-0026003 (ultrasound measurements and data analysis). Research at the University of Maryland was supported by
	the Gordon and Betty Moore Foundation’s EPiQS Initiative Grant No. GBMF9071 (materials synthesis), the Department of Energy, Office of Basic Energy Sciences Award No. DE-SC-0019154 (sample characterization), the NIST Center for Neutron Research, and the Maryland Quantum Materials Center. Research at the University of Wisconsin was supported by the Simons Foundation under Grant No. SFI-MPS-NFS-00006741-02 (intra-band theory). Collaborative exchange was funded in part by a QuantEmX grant from ICAM and the Gordon and Betty Moore Foundation through Grant GBMF9616.
	
	\section{Data Availability Statement}
	The data that support the findings of this article are openly available at \url{https://github.com/CHiLL-Ramshaw/manuscripts-supporting_data/tree/2f152da924c294dfc4c06ac37db01ddcde2f2905/2026_UTe2_Vanishing}.
	
	\section{Appendix I: Transverse-mode $c_{55}$ data}
	
	\begin{figure*}[t]
		\begin{center}
			\includegraphics[width=\textwidth]{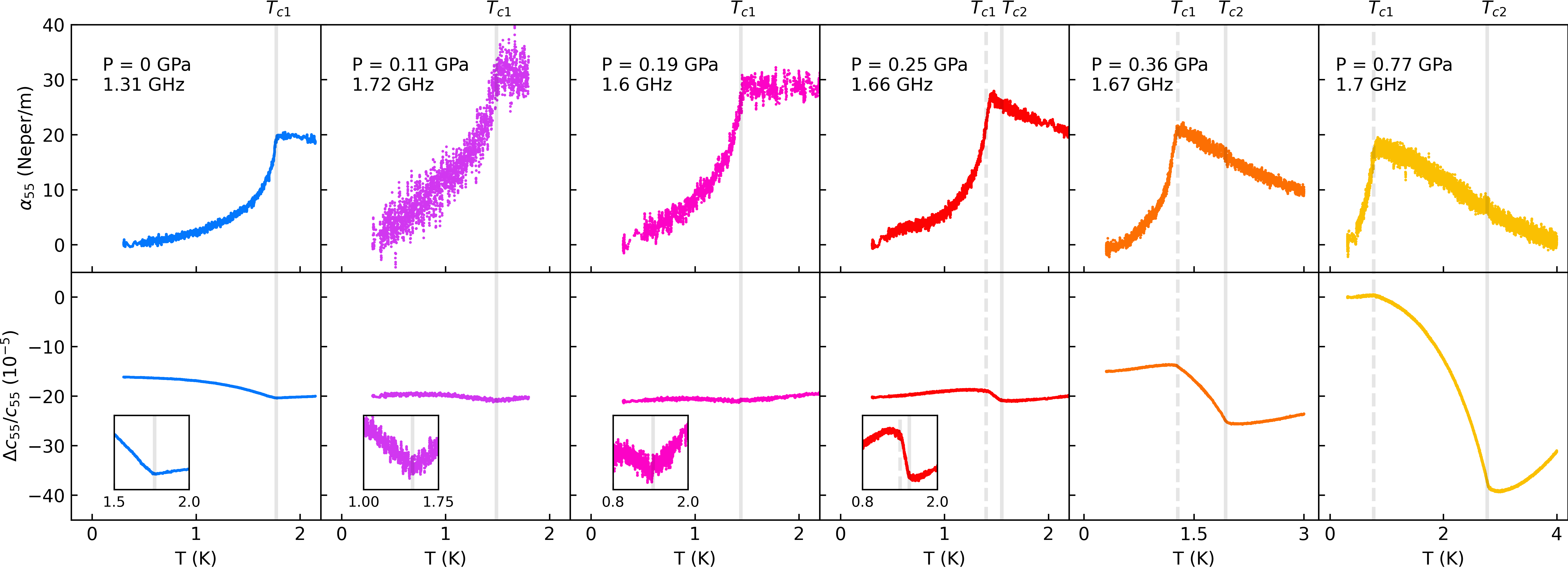}
		\end{center}
		\caption{Sound attenuation and the shear elastic modulus $c_{55}$. There is a sharp drop in the attenuation at \Tco at all pressures, following the conventional expectation of BCS theory. Sharp kinks are seen in the elastic modulus at both \Tco and \Tct. Insets show a zoomed-in elastic modulus, clearly showing two \Tcs for $P>\Ps$.}
		\label{fig:c55panels}
	\end{figure*}
	
	In addition to the \ctt compressional elastic modulus shown in the main text, we have also measured a $c_{55}$ shear modulus. Data at 0.11 GPa and 0.19 GPa is measured on sample 1, 0.25 GPa to 0.77 GPa is measured on sample 2, and the 0 GPa data is sample 3. This data is shown in \autoref{fig:c55panels}. For a one-component order parameter we do not expect either a jump at \Tc in the shear modulus \cite{theussSinglecomponentSuperconductivityUTe22024a} nor a Landau-Khalatnikov peak in the shear attenuation \cite{miyakeLandauKhalatnikovDampingUltrasound1986}. This is exactly what we see in the data of \autoref{fig:c55panels}, with a sharp drop in the attenuation at \Tco and sharp kinks in the elastic modulus at both \Tco and \Tct (for pressures $P>\Ps$). The sharp features at both \Tcs indicate minimal pressure broadening over the entire pressure-temperature space we measure. 
	
\begin{figure}[h]
		\centering
		\includegraphics[width=\columnwidth]{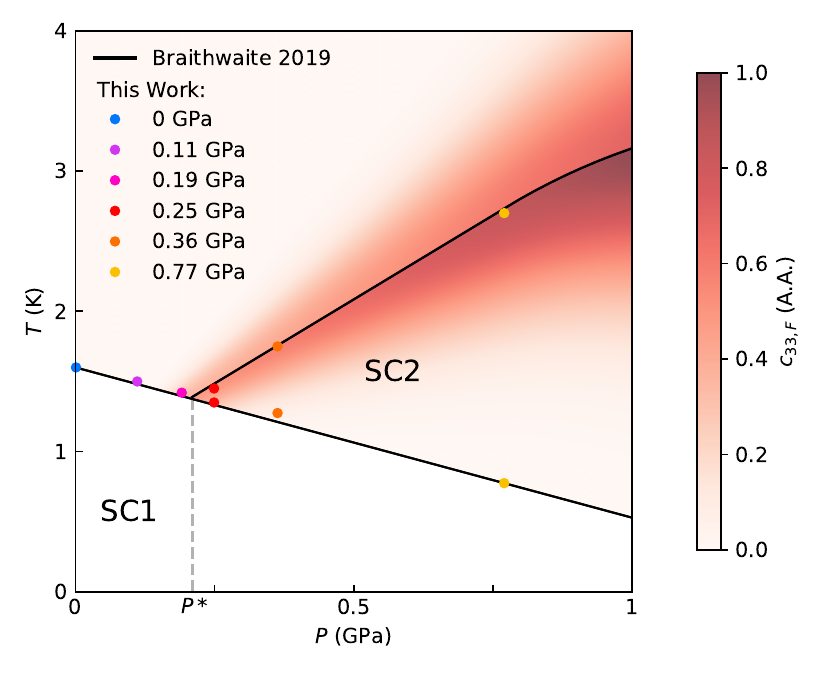}
		\caption{A schematic diagram of the fluctuation contribution ($c_{33,F}$) to \ctt overlaid on the \ute phase diagram. Deeper red color indicates a larger fluctuation contribution. Colored triangles indicate \Tcs at the pressures at which we measure. Due to the absence of a phase boundary in the normal state, any thermodynamic variable (like $c_{33,F}$) cannot change discontinuously as pressure is varied in the normal state. This causes a ``spillover'' of the fluctuation contribution to pressures less than \Ps, leading us to observe fluctuations at $P=0.19$ GPa at temperatures just above \Tc.}
		\label{fig:colorplot}
	\end{figure}

	The $c_{55}$ data in \autoref{fig:c55panels} show that we measure a single transition at $P=0.19$ GPa, indicating that the sample only enters the SC1 phase at this pressure. We attribute the fluctuations we observe in the \ctt data at this pressure to its proximity to the SC2 phase. We illustrate this by showing a schematic of the (diverging) fluctuation contribution $c_{33,F}$ to \ctt as a colorplot on the $P-T$ phase diagram (\autoref{fig:colorplot}):  
	\begin{equation}
		\Delta c_{33,F} = 
		-\left( \frac{d T_c}{d \epsilon_{zz}} \right)^2 \frac{\Delta C_{\text{MF}}}{T_C} \times 
		\begin{cases}
			\sqrt{\frac{Gi}{2t}} & T>\Tc \\
			\sqrt{\frac{Gi}{t}} & T<\Tc.
		\end{cases}
	\end{equation}

	\section{Appendix II: Landau-Khalatnikov Damping}
	\label{sec:lkdamping}
	
	\begin{figure*}[t]
		\begin{center}
			\includegraphics[width=0.8\textwidth]{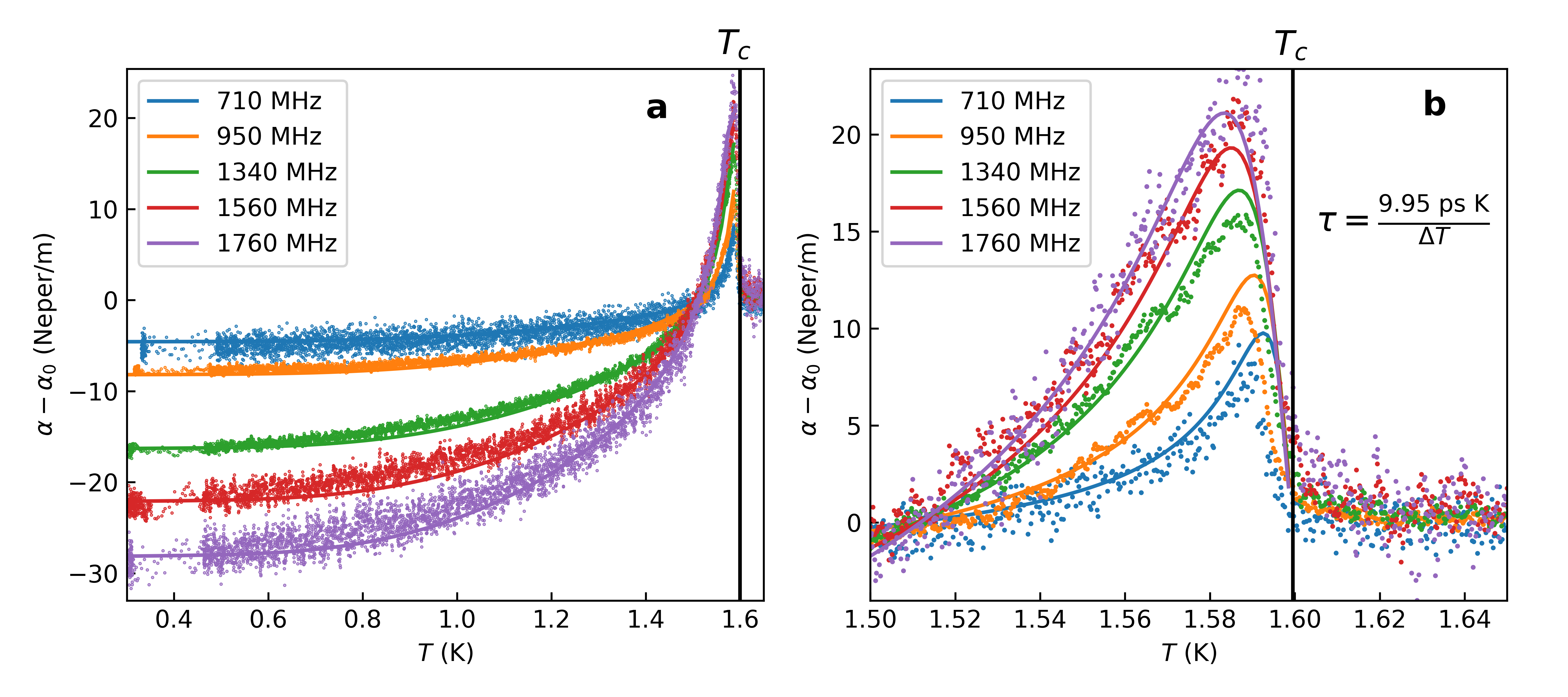}
		\end{center}
		\caption{\ctt sound attenuation at zero pressure at different frequencies, along with fits to the model described in the Landau-Khalatnikov Damping section. a) Data from 0.3 to 1.65 K. Solid lines correspond to the fits. b) Data closer to \Tc, showing the LK peak in greater detail. The fit described by \autoref{eq:peak} models the data very well, with just three fit parameters used to fit all the data we measure.}
		\label{fig:LKfits}
	\end{figure*}
	
	In the main text, we attribute the peaks observed in \att at \Tc to order parameter relaxation. Here, we construct a mean-field model of order parameter relaxation and fit it to the 0 GPa data---where we have a broad frequency depenedence of \att---to show good quantitative agreement. 
	
	We begin by considering a Ginzburg-Landau free energy density ($f$) expression in the presence of a strain $\epsilon$. The coupling between the strain and the superconducting order parameter $\psi$ is proportional to $\epsilon |\psi|^2$, which is the lowest order coupling between a strain and a superconducting order parameter that preserves gauge symmetry \cite{theussSinglecomponentSuperconductivityUTe22024a}.
	
	\begin{equation}
		f =  f_n + \alpha\left(T\right) \left| \psi \right|^2 + \frac{1}{2}\beta \left| \psi \right|^4 - \gamma \epsilon|\psi|^2
	\end{equation}
	
	$\alpha(T) \equiv \alpha_0(T - \Tc)$ and $\beta$ are the Ginzburg-Landau phenomenological parameters that model the symmetry breaking phase transition. $\gamma$ sets the strength of the strain-order parameter coupling.
	
	This coupling is only allowed when the product $\epsilon |\psi|^2$ is a scalar, and for an orthorhombic system like \ute, can only happen for a longitudinal strain, or for a shear strain with a two component order parameter (ruled out for the SC1 phase by \citet{theussSinglecomponentSuperconductivityUTe22024a}).
	
	In the mean field approximation, we can find the equilibrium value of the order parameter magnitude $|\psi_0|$ for a given strain and temperature by setting $\partial f/ \partial \psi = 0$ and solving for $\psi$. If the order parameter is shifted away from this equilibrium value (e.g. by the application of strain), we expect it to return back to equilibrium. It is here that we make our first assumption about the dynamics of the system. A deviation of the order parameter magnitude $|\psi| = |\psi_0| + |\delta \psi|$ from equilibrium decreases as
	\begin{equation}
		\frac{\partial(|\delta \psi|)}{\partial t} = -\xi \frac{\partial f}{\partial |\psi|}, 
		\label{eq:secondlaw}
	\end{equation}
	where $\xi$ quantifies the ``inertia'' of the order parameter to fluctuations. It is instructive to view this equation as an analogue of Newton's second law of motion, $a = \frac{1}{m} F$, with $\phi = \frac{\partial f}{\partial |\psi|}$ playing the role of a generalized force that pulls $\psi$ back to equilibrium. We account for fluctuations only in the \textit{magnitude} of the order parameter, since there is no energy cost associated with (mean-field) \textit{phase} fluctuations in the superconducting state.
	
	In our pulse-echo experiment, fluctuations $\delta \psi$ are caused by the oscillating strain field $\epsilon$. For small strains, and hence small fluctuations, there is a small restoring force $\delta \phi = \frac{\partial \phi}{\partial |\psi|} |\delta \psi| +  \frac{\partial \phi}{\partial \epsilon} \delta \epsilon$. We define $Z \equiv \partial \phi/\partial \epsilon = \partial^2 f/ \partial \epsilon \partial |\psi|$ and  $Y \equiv \partial \phi/\partial |\psi| = \partial^2 f/\partial |\psi|^2$ to obtain an equation of motion for small order parameter fluctuations $\delta \psi$ induced by a time varying strain $\delta \epsilon$:
	\begin{equation}
		\frac{\partial|\delta \psi|}{\partial t} = -\tau^{-1} |\delta \psi| -\xi Z \delta \epsilon.
		\label{eq:eqofmotion}
	\end{equation}
	Here $\tau^{-1} = \xi Y$ is the damping rate for order parameter fluctuations, and the second term can be thought of as a forcing function. 
	
	For an oscillating strain, we make the ansatz $\delta \epsilon = \epsilon_0 e^{-i \omega t}$ and $|\delta \psi| = |\delta \psi| (\omega) e^{-i 
		\omega t}$. The equation of motion then gives
	\begin{equation}
		|\delta \psi (\omega)| = \frac{Z \epsilon_0}{Y} \frac{1}{i\omega \tau  - 1}.
		\label{eq:opsolution}
	\end{equation}
	We can then use $\sigma = \partial f/\partial \epsilon$ to find the stress response to the strain, $\sigma = \sigma_0e^{-i\omega t}$. The work done (and hence energy lost) per unit time is given by $\omega \text{ Im}(\sigma_0) \text{Re}(\epsilon_0) $, which gives for the attenuation
	\begin{equation}
		\alpha_{\text{LK}} = \frac{\omega^2 \tau^2}{1 + \omega^2 \tau^2} 4 \xi \gamma^2 |\psi_0|^2 \epsilon_0^2.
		\label{eq:peak}
	\end{equation}
	The only two temperature dependent terms in this equation are the timescale $\tau$ and the equilibrium order parameter $\psi_0$. In the small strain approximation these become $\tau = \frac{1}{\xi Y} \approx \frac{1}{\xi \alpha_0 (T_c - T)}$ and $|\psi_0| \approx \sqrt{\frac{\alpha_0(T_c - T)}{\beta}}$. $\tau$ diverges as the system approaches $T_c$ \cite{miyakeLandauKhalatnikovDampingUltrasound1986}, leading to a large attenuation near $T_c$. The order parameter $\psi_0$ itself, however, is zero at $T_c$, driving the attenuation to zero at $T_c$. The combination of these two factors leads to a peak in the attenuation slightly below $T_c$. 
	
	Using the temperature dependence for $\tau$ and $|\psi_0|$, we solve $\partial \alpha_{LK}/\partial T = 0$ to obtain the position of the attenuation peak. This gives us $T_c - T_{\text{peak}} = \Delta T_{\text{peak}} = \omega/2\alpha_0\xi$. We insert this value back in \autoref{eq:peak} to find the peak attenuation:
	\begin{equation}
		\alpha_{\text{peak}} = \omega \frac{\gamma^2}{ \beta} \epsilon_0^2 = \omega \Delta c \epsilon_0^2
		\label{eq:linearinfreq}
	\end{equation}
	$\alpha_{\text{peak}}$ is linear in frequency with a magnitude that only depends on the size of the elastic modulus jump $\Delta c$. We add to the attenuation from \autoref{eq:peak} a gap attenuation $\alpha_{\text{gap}} = \alpha_N f(\Delta (T))$ where $\alpha_N$ is the normal state attenuation, $f$ is the Fermi-Dirac distribution function, and $\Delta(T)$ is the magnitude of the superconducting gap \cite{khanSoundAttenuationElectrons1987} to obtain a total attenuation. We fit this total attenuation to the data in \autoref{fig:LKfits}.
	
	\section{Appendix III: Origin of dominant inter-band pairing}
	
	To understand the origin of a dominant inter-band pairing and the corresponding reduced phase stiffness, we consider a simple model motivated by ferromagnetic spin-fluctuations in UTe$_2$. A key justification for this model is that DFT calculations find that the dominant magnetic interaction is a ferromagnetic interaction between the $f$-electrons on the two U atoms within the unit cell \cite{xuQuasiTwoDimensionalFermiSurfaces2019}. These ferromagnet correlations are consistent both with neutron observations of spin fluctuations at ambient pressure \cite{Knafomagneticfluctuations2021} and with the magnetic order observed under pressure \cite{Knafomagneticorder2025}. As shown in more detail below, these ferromagnetic interactions drive spin-triplet superconducting pairing \cite{shishidouTopologicalBandSuperconductivity2021} but with a pairing interaction that is largest for inter-band pairing. As recently shown by Samokhin \cite{samokhinGinzburgLandauEnergyMultiband2024}, this dominant inter-band pairing regime allows a dominant inter-band superconducting state with a finite $T_c$ that is driven by a Josephson-like coupling to more usual intra-band pairing. Importantly, as shown in Ref.~\cite{samokhinGinzburgLandauEnergyMultiband2024} and described in more detail below, this inter-band pairing state is much more susceptible to phase fluctuations than the more usual intra-band Cooper pairs.   
	
	More specifically, we consider a model for itinerant $f$-electrons on the two U sites within the unit cell. For clarity, we do not include the Te electrons and also do not include spin-orbit coupling (these will not change the qualitative arguments). The single-particle Hamiltonian that describes these fermions is 
	\begin{equation}
		H=\epsilon_0(k) +f_x(k)\tau_x+f_y(k)\tau_y,
		\label{eq:H0}
	\end{equation}
	where the $\tau$ operators are Pauli matrices that operate in sublattice space. We take $f_x(k)=t_1$ and $f_y(k)=t_2\sin \frac{k_z}{2}\cos \frac{k_x}{2} \cos \frac{k_y}{2}$. An important parameter in what follows is the ratio  $\epsilon=t_2/t_1$. This parameter dictates the relative size of inter-band pairing to intra-band pairing interactions. If it is small, the inter-band pairing interaction dominates. Fits to the observed Fermi surface at ambient pressure give $\epsilon<<1$ ($\epsilon\approx 6\times 10^{-3}$) \cite{theussAbsenceBulkThermodynamic2024a} and it is physically reasonable that such an inequality also applies at the pressures relevant $P\approx P^*$ here since $t_1$ is a nearest neighbor hopping and $t_2$ is a longer range hopping.  
	
	For these $f$-electron fermions, the intra-unit cell ferromagnetic interaction is given by
	\begin{equation}
		H_{M}=-|J|\sum_i \vec{S}_{i,1} \cdot \vec{S}_{i,2},
	\end{equation}
	where $\vec{S}_{i,j}=c^{\dagger}_{i,j,s}\vec{\sigma}_{s,s^\prime} c_{i,j,s^{\prime}}$
	where $c^{\dagger}_{i,j,s}$ creates a $f$-electron in unit cell $i$, U position $j$ within the unit cell, and spin $s$. This leads to the following pairing interaction:
	\begin{equation}
		S_{C}=-\frac{|J|}{V}\int_{k,p,j}\left(\psi^\dagger_{k}\tau_y \vec{ \sigma}_j i\sigma_y \psi^*_{-k}\right)\left(\psi^\text{T}_{-p}\tau_y \vec{\sigma}_j i\sigma_y \psi_{p}\right)\label{eq:SIC}
	\end{equation}
	where $\psi_{k}=(c_{1,{\bf k},\uparrow}, c_{1,{\bf k},\downarrow},c_{2,{\bf k},\uparrow},  c_{2,{\bf k},\downarrow})^T$ and $V$ is volume. This pairing interaction implies that a spin-triplet sublattice-singlet pairing state, characterized by $\tau_y \sigma_i (i\sigma_y)$, is stabilized by these ferromagnetic fluctuations \cite{shishidouTopologicalBandSuperconductivity2021}.  
	While describing the pairing as a spin-triplet sublattice-singlet provides a convenient real-space description of superconducting pairing, superconductivity is more naturally described in band (or momentum) space. In general, any real-space Cooper pairing will decompose into usual intra-band Cooper pairs together with inter-band Cooper pairs. Typically, it is assumed that the relevant physics is captured by the intra-band Cooper pairing, for which a usual weak-coupling superconducting state naturally emerges \cite{sigristPhenomenologicalTheoryUnconventional1991}. However, this implicitly assumes that the inter-band pairing interaction can be largely neglected. As pointed out by Samokhin \cite{samokhinGinzburgLandauEnergyMultiband2024}, when the inter-band pairing interaction is sufficiently large, a dominant inter-band pairing state can emerge. This state is in part stabilized by a Josephson-like coupling between the inter-band and intra-band Cooper pairs. Such a Josephson-like coupling appears because the inter- and intra-band pairing channels all share the same symmetry. 
	
	To see how a predominantly interband pairing state can emerge from Eqs.~\ref{eq:H0} and \ref{eq:SIC}, we express Eq.~\ref{eq:SIC} in the band basis. The eigenstates in the band basis are found by applying the following unitary transformation to \ref{eq:H0}
	\begin{equation}
		U_k=\frac{1}{\sqrt{2}}\begin{bmatrix}
			1 & - e^{-i\phi_k} \\ 
			e^{i\phi_k}& 1 
		\end{bmatrix}
	\end{equation}
	where $e^{i\phi_k}=[f_x(k)+if_y(k)]/f(k)$ and $f(k)=\sqrt{f_x^2(k)+f_y^2(k)}$ .
	The consequence of this transformation is to convert the leftmost operator $\tau_y$ in Eq.~\ref{eq:SIC} to 
	\begin{equation}
		U_k\tau_yU^T_{-k}=\begin{bmatrix}
			-\sin \phi_k  & - ie^{-i\phi_k}\cos \phi_k \\ 
			i e^{i\phi_k}\cos\phi_k& \sin\phi_k\label{eq:SCband}
		\end{bmatrix}.
	\end{equation}
	In the above expression, the diagonal elements describe the usual intra-band Cooper pairs while the off-diagonal elements describe the inter-band pairing. Since $\sin\phi_k\approx \epsilon \cos\phi_k$, this implies that the interaction for the inter-band pairing is parametrically larger than that for the intra-band pairing. Following Ref.~\cite{samokhinGinzburgLandauEnergyMultiband2024},  Eq.~\ref{eq:SCband} can be re-expressed explicitly in the band basis as
	\begin{eqnarray}
		S_C=&&\frac{1}{2V}\sum_{k,k',q}\sum_{n_i,s_i}V^{n_1n_2n_3n_4}_{s_1s_2s_3s_4}(k,k')\nonumber \\&&\times a^{\dagger}_{k+\frac{q}{2},n_1,s_1}\tilde{a}^{\dagger}_{k-\frac{q}{2},n_2,s_2}\tilde{a}_{k'-\frac{q}{2},n_3s_3}a_{k'+\frac{q}{2},n_4s_4} \nonumber \\&&
	\end{eqnarray} 
	where $\tilde{a}^{dagger}_{k,ns}=Ka^{dagger}_{k,ns}K^{-1}$ is time reversal ($K$) partner of $a^{dagger}_{k,ns}$.   We find that the intra-band gap functions are usual odd-parity spin-triplet state with a momentum dependence proportional to $\sin \frac{k_z}{2}\cos \frac{k_x}{2} \cos \frac{k_y}{2}$. Furthermore, the inter-band gap function is also spin-triplet, is antisymmetric under the interchange of the two bands, and predominantly momentum independent. Since the symmetry of all three order parameters is odd-parity, it is natural to ask how this appears in the inter-band pairing state (which is momentum independent). This can be understood from Eq.~\ref{eq:H0} in the limit $t_2=0$. In this limit, the two bands that emerge are formed from bonding and anti-bonding states of the two U atoms. Since the parity symmetry operation exchanges the two U atoms in the unit cell, these bonding and anti-bonding states are of opposite parity. Consequently, the momentum-independent inter-band order is odd-parity since it pairs fermion states with opposite parity.
	
	To understand the hierarchy of the transition temperatures of the possible pairing states, it is useful to find  effective interactions $V^{n_1n_2n_3n_4}$ in powers of $\epsilon$. In our case, we find that the intra-band pairing  (with order parameters $\psi_1$ and $\psi_2$) $V^{iijj}\propto |J|\epsilon_2$, that the inter-band pairing (with order parameter $\psi_m$) terms $V_{1212}\propto |J|$, and that the Josephson coupling between the intra-band and inter-band pairs $V^{ii12}\propto |J|\epsilon$. Following Ref.~\cite{samokhinGinzburgLandauEnergyMultiband2024}, this implies that the usual intra-band pairing, for which $(\vec{\psi}_1,\vec{\psi}_2,\vec{\psi}_m)\approx (1,1,\epsilon)$,  and the dominant inter-band pairing state, for which $(\vec{\psi}_1,\vec{\psi}_2,\vec{\psi}_m)\approx (\epsilon,\epsilon,1)$ both have a $T_c$ that is of order $\epsilon^2$. The pairing state that is stabilized then depends upon additional details of the relevant electronic states. At ambient pressure, where there is still some debate about the essential fermiology \cite{liu2024densityfunctionaltheorybased}, this appears to be reasonably well understood, and  the pairing state (without any magnetic fields applied) is consistent with a usual intra-band pairing state.   However, for pressures $P\approx P^*$,  the essential fermiology is much less well understood. Here we suggest that the dominant inter-band pairing state has the highest $T_c$. 
	
	To determine the phase stiffness, the Ginzburg Free energy to quadratic order must be found. It is possible to do so using standard techniques  \cite{samokhinGinzburgLandauEnergyMultiband2024}, with the result 
	\begin{equation}
		F_2=\vec{\psi}^\dagger \hat{A} \vec{\psi}+  \kappa_1|\vec{D}\psi_1|^2+\kappa_2|\vec{D}\psi_2|^2+\tilde{\kappa}|\vec{D}\psi_m|^2
	\end{equation}
	where $\hat{A}$ describes the order parameter coupling originating from $V_{n_1n_2n_3n_4}$ (the detailed form of $\hat{A}$ is given in Ref.~\cite{samokhinGinzburgLandauEnergyMultiband2024}), $\vec{\eta}=(\psi_1,\psi_2,\psi_m)^T$, $\kappa_i$ are the usual coefficients for intra-band Cooper pairs (below denoted a $\kappa_{usual}$) and $\tilde{\kappa}$ is the corresponding coefficient for inter-band Cooper pairs. Samokhin showed that $\tilde{\kappa}/\kappa_i\approx T_c^2/t^2<<1$.
	Restricting ourselves to the interband pairing state (which here is the eigenstate of $\hat{A}$ with the largest $T_c$),  $(\psi_1,\psi_2,\psi_m)\approx \psi(\epsilon,\epsilon,1)$, then the  effective coefficient for this state is given by 
	\begin{equation}
		\kappa_{eff}\approx \epsilon^2 \kappa_{usual}+\tilde{\kappa}<<\kappa_{usual}   
	\end{equation}
	revealing that the dominant interband pairing state is much more susceptible to fluctuations. Note that we have assumed the $T_c$ for the predominantly intra-band pairing states are much lower than that for the predominantly inter-band pairing state, so that their contribution to the phase stiffness can be ignored.

	\section{Appendix IV: Ginzburg number at ambient pressure}
	
	At ambient pressure, where the broadening due to fluctuations is negligible, there is still a contributiton to the width of \Tc from impurity and/or inhomogeneous strain broadening. Thus we cannot use the width of the transition alone as a reliable estimate of the Ginzburg number at ambient pressure. In this section, we estimate $Gi$ from ambient pressure coherence length, penetration depth, and specific heat data. We make three independent estimates of $Gi$ and show that they agree to within an order of magnitude. Note that these rough estimates of $Gi$ are only intended as a point of comparison between the SC1 and SC2 phases.
	
	From the penetration depth data of Ishihara et. al. \cite{ishiharaChiralSuperconductivityUTe22023} we have
	\begin{equation}
		\begin{aligned}
			\lambda_a\left(0\right) &= 1420 \si{ \nano \meter} \\
			\lambda_b\left(0\right) &= 710 \si{ \nano \meter} \\
			\lambda_c\left(0\right) &= 2750 \si{ \nano \meter}. 
		\end{aligned}
		\label{eq:lambda}
	\end{equation}
	From the STM data of Sharma et. al. \cite{sharmaObservationPersistentZero2025}, the $a$-axis and $bc$ plane coherence lengths are found to be
	\begin{equation}
		\begin{aligned}
			\xi_a\left(0\right) &= 12\si{ \nano \meter} \\
			\xi_{bc}\left(0\right) &= 4\si{ \nano \meter}.
		\end{aligned}
		\label{eq:xi}
	\end{equation}
	
	We make another independent estimate of the coherence length using the upper critical field \cite{vasinaQuantitativeThermodynamicStudy2026} \cite{ramshawVortexLatticeMelting2012}:
	\begin{equation}
		\begin{aligned}
			\xi(0) &= \sqrt{\frac{\Phi_0}{2\pi \mu_0 H_{c2}(T = 0)}} \\
			&= 4\si{\nano \meter}.
		\end{aligned}
		\label{eq:brad}
	\end{equation}
	
	First, we use the formula from Brandt et. al. \cite{brandtFluxlineLatticeSuperconductors1995} to estimate the Ginzburg number from the zero temperature penetration depth and coherence length:
	\begin{equation}
		Gi = \frac{1}{2} \left( \frac{k_B \Tc \mu_0}{4 \pi \xi(0)^3 B_c(0)^2} \right)^2
		\label{eq:gi_lambda}
	\end{equation}
	Here $B_c$ is the thermodynamic critical field, defined as
	\begin{equation}
		B_c = \frac{\phi_0}{2\sqrt{2}\pi\lambda\xi}
	\end{equation}
	Since the $\lambda(0)$ and $\xi(0)$ of \autoref{eq:xi} and \autoref{eq:lambda} are anisotropic, we use their geometric mean to estimate $Gi$. This is motivated by how $Gi$ is derived by solving the fluctuation contributions from Ginzburg Landau theory \cite{larkinTheoryFluctuationsSuperconductors2005}. $\xi(0)$ as derived from \autoref{eq:brad} is of the same order of magnitude and so does not change our conclusions.  We get
	\begin{equation}
		Gi(\mathrm{from \ \ensuremath{\lambda} \ and \ \ensuremath{\xi}}) = 4.6 \times 10^{-5}
	\end{equation}
	The Ginzburg number can also be expressed in terms of the specific heat jump at \Tc and the coherence length, which can be shown to be precisely equal to the Ginzburg number of  \autoref{eq:gi_lambda} \cite{larkinTheoryFluctuationsSuperconductors2005}: 
	\begin{equation}
		Gi = \frac{1}{2} \left( \frac{ 2 k_B}{\Delta C \xi^3} \right)^2
	\end{equation}
	Using the specific heat data of \cite{ranNearlyFerromagneticSpintriplet2019}, we estimate $\Delta C/T = 0.2\ \si{\joule \per \mole \kelvin \squared}$. This gives for the Ginzburg number
	\begin{equation}
		Gi(\mathrm{from \ \ensuremath{\Delta C/T} \ and \ \ensuremath{\xi}}) = 0.8 \times 10^{-5}
	\end{equation}
	
	\section{Appendix V: The Kinetic Inductance of \ute}
	
	Kinetic inductance detectors (KIDs) are highly sensitive single-photon detectors widely used in astrophysical applications. These devices rely on the kinetic inductance of a superconducting film, which sets their sensitivity and scales inversely with the superfluid phase stiffness. In conventional KID materials, such as granular aluminum and NbTiN, low phase stiffness is achieved by introducing disorder into an otherwise stiff superconductor, suppressing the superfluid fraction and enhancing the kinetic inductance. These high kinetic inductance superconductors are used for small-footprint, high-impedance qubits. 
	
	In this section, we compute the kinetic inductance of a thin film of \ute\ in its pressure-induced SC2 state, which could potentially be stabilized outside of a pressure cell using epitaxial strain. We find that \ute in the SC2 state has a kinetic inductance that is comparable to that of state-of-the-art disordered $s$-wave superconductors, but in this case it arises intrinsically, without the need for disorder.
	
	We use the known value for the heat capacity jump \cite{vasinaConnectingHighFieldHighPressure2025} and the value we calculate for $Gi$ at $P = 0.77$ GPa in \autoref{eq:ginzburg} to obtain the phase stiffness $\kappa_0$ as
	\begin{equation}
		\kappa_0 = 1.0 \times 10^{-14}\ \si{\joule/\meter}.
	\end{equation}
	A thin film made of a superconductor with a phase stiffness $\kappa_0$ has a kinetic inductance $L_K$ given by \cite{kreidelMeasuringKineticInductance2024}
	\begin{equation}
		L_K = \frac{1}{8} \left(\frac{\hbar^2}{e^2 \kappa_0^2}\right)\left(\frac{1}{t}\right),
	\end{equation}
	where $\hbar$ is the reduced Planck's constant, $e$ is the electronic charge, and $t$ is the thickness of the superconducting film. For a 3 nm thick \ute film stabilized in its SC2 state with $Gi=0.6$ we obtain a kinetic inductance of 1.8 nH/$\square$. This intrinsic kinetic inductance is comparable to the high kinetic inductance materials used today \cite{bretz-sullivanHighKineticInductance2022}, and is obtained without the need for any disorder.
	
	\section{Appendix VI: Comparison with specific heat}
	
\begin{figure}[h]
		\begin{center}
			\includegraphics[width=0.5\textwidth]{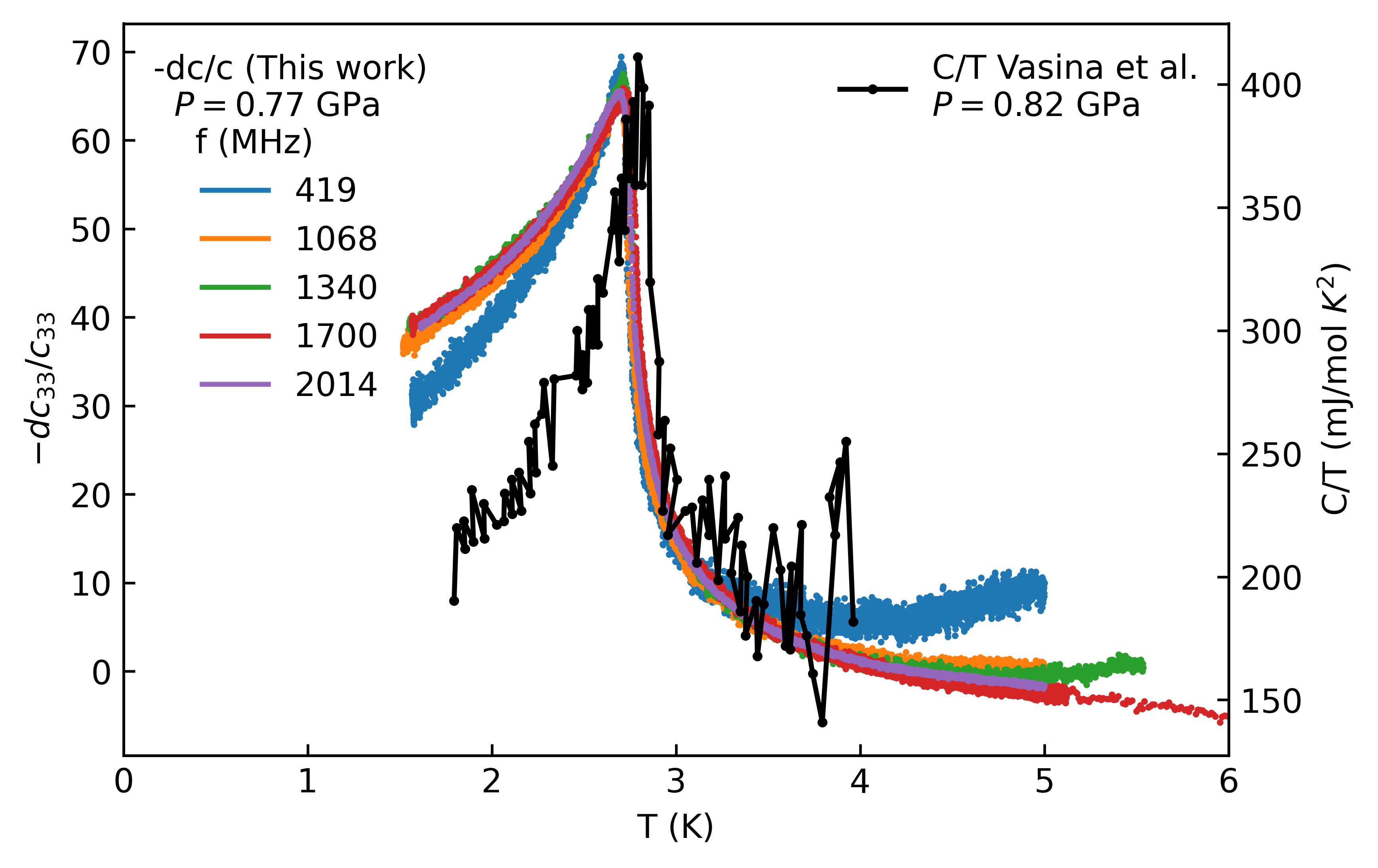}
		\end{center}
		\caption{\textbf{Fluctuations in the specific heat and \ctt elastic modulus}. We show \ctt measurements at different frequencies (on an inverted scale) at 0.77 GPa plotted with specific heat ($C/T$) data from Vasina et al. at 0.82 GPa \cite{vasinaQuantitativeThermodynamicStudy2026}. Fluctuations in both the data are consistent, as is expected of critical fluctuations.}
		\label{fig:heatcap}
	\end{figure}

	A central point of this work is that the critical fluctuations we observe should also be visible in other thermodynamic quantities such as the specific heat. In \autoref{fig:heatcap}, we plot our data against recent quantitative specific heat measurements \cite{vasinaQuantitativeThermodynamicStudy2026}. The data overlap well for $T>\Tc$, showing consistent fluctuations. For $T<\Tc$ the strain-order parameter coupling causes the data to deviate (i.e. the shape below \Tc is not universal).
	
	If the broadening at \Tct were extrinsic, both measurements would have to experience \textit{exactly} the same pressure distribution, both quantitatively and qualitatively. This is extremely unlikely: the sample sizes used for ultrasound ($<$0.6 mg) and quantitative heat capacity (23.73 mg \cite{vasinaQuantitativeThermodynamicStudy2026}) are vastly different, and so are the pressure media (Methanol-Ethanol and Fluorinert \cite{vasinaQuantitativeThermodynamicStudy2026} respectively). On the other hand, the overlap is exactly what is expected if the fluctuations were caused by thermodynamic criticality.
	
%
%

\clearpage
\onecolumngrid

\begin{center}
	{\LARGE\bfseries Supplementary Information}
\end{center}

\vspace{1em}

\section{Pressure Homogeniety}
\label{supplement}

\begin{figure}[h]
	\centering
	\includegraphics[width=0.8\textwidth]{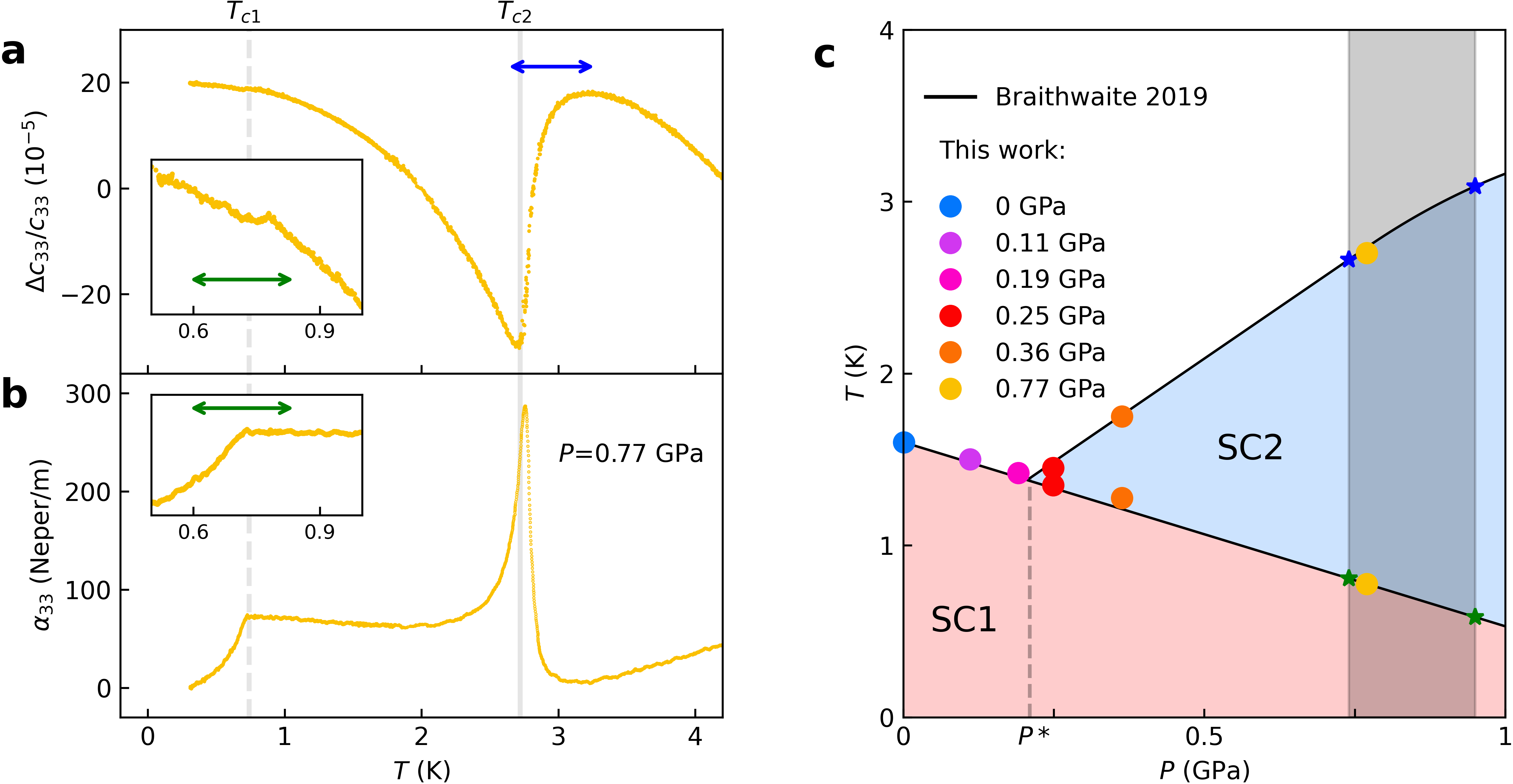}
	\caption{\textbf{A quantitative analysis of pressure inhomogeniety}. The blue arrow in (a) represents the width of the SC2 transition we see at 0.77 GPa --- the data with the largest fluctuation region. The pressure inhomogeniety required to produce such a broadening is represented by the gray bar in the phase diagram of (c), with blue stars representing the upper and lower extent of the transition. This inhomogeniety would lead to a transition bounded by the green stars on the SC1 phase boundary. When plotted against the data as the green arrow in the insets of (a) and (b), the true width of the SC1 transition is seen to be much narrower (in particular, it is easy to see that there is no precursor decrease in $\alpha_{33}$ above \Tct), implying that the broadening at SC2 is an intrinsic effect.}
	\label{fig:broadening}
\end{figure}

In this section, we quantitatively analyze the broadening observed at \Tct for $P = 0.77$ GPa to show that it is an intrinsic effect that cannot be explained by pressure inhomogeniety. We indicate the apparent ``width" of \Tct with a blue arrow in \autoref{fig:broadening}a, extending from $T=2.7$ K to $T= 3.1$ K. If we assumed that the ``intrinsic" \Tc was sharp, then the pressure inhomogeneity required to produce such a width in \Tct is $\Delta P= 0.21$ GPa, which we indicate with a gray bar in \autoref{fig:broadening}c, with \Tct bounded by the blue stars. Such a spread in pressure would correspond to a spread of \Tc's at \Tco from $T=0.6$ K to $T=0.8$ K---indicated by green stars in \autoref{fig:broadening}c. We represent this expected spread as a green arrow in the insets of \autoref{fig:broadening}a and b, which show a zoomed in view of the \Tco transition. 
Experimentally, however, we measure a much sharper transition ($\Delta T= 30$ mK), indicating minimal pressure broadening $\Delta P<$ 0.03 GPa.

\section{Background subtraction}

\begin{figure}[h]
	\centering
	\includegraphics[width=0.8\textwidth]{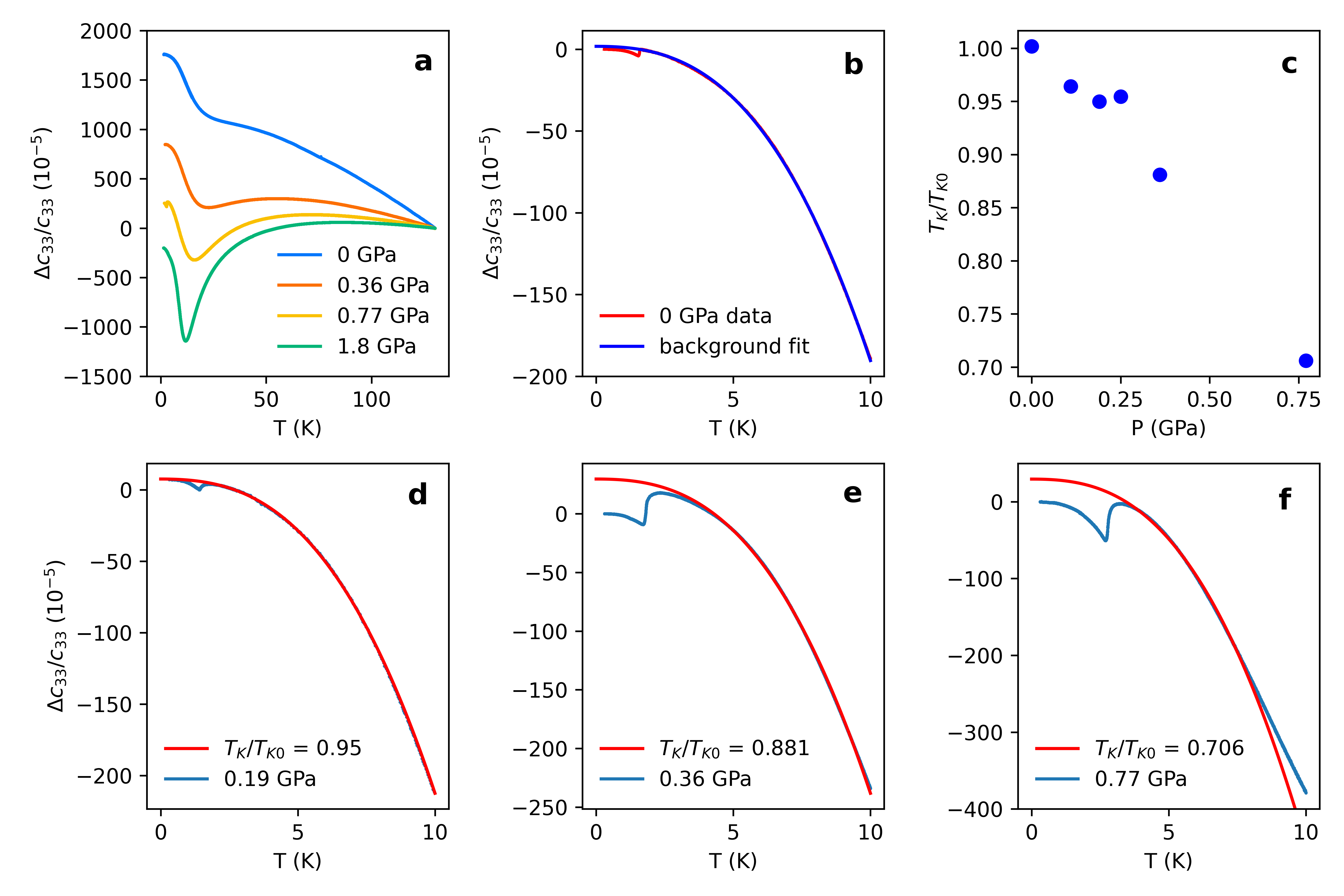}
	\caption{Background subtraction procedure. a) Raw $c_{33}$ elastic modulus from $T=2$ to 130 K. b) A polynomial fit to the 0 GPa data for $T>\Tc$. c) A plot of the Kondo temperature (extracted from background fits) as a function of pressure. d,e,f) Background fits at three representative pressures. A single variable --- the Kondo temperature --- is used as a fitting parameter. A deviation of the fit from the data is observed at 0.77 GPa for $T>8$ K due to the proximity of the minima in \ctt to the fitting range (the mimimum is clearly observable near $\approx  20$ K in panel a). }
	\label{fig:bgsub}
\end{figure}

\autoref{fig:bgsub}a shows the raw \ctt data as a function of temperature up to 130 K. These data resemble what is seen for other heavy fermion compounds \cite{nakamura_elastic_1991,pan_ultrasonic_2023} and are influenced by two factors---a conventional lattice stiffening that monotonically increases \ctt down to low temperatures, and a Kondo softening that produces the characteristic ``dip'' at $\approx$20 K, which can be seen most clearly in the 1.8 GPa data. We observe two changes in these curves as a function of increasing pressure: first, the Kondo effect contribution to the elastic constants grows, and second, the temperature at which the minimum occurs decreases.

At temperatures less than 10 K we expect the lattice hardening to have flattened out \cite{varshni_temperature_1970}, and so the primary energy scale that governs the background is the Kondo energy scale $T_K$. In \autoref{fig:bgsub}b, we fit the \ctt data at zero pressure to obtain a background. We don't observe any precursor fluctuations at this pressure, so an accurate background is obtained by fitting a polynomial $f(T)$ to the data from just above $T_C$ up to 10 K. 

As the data is governed by a single pressure-dependent energy scale $T_K$, the behavior at higher pressures can be captured by a simple rescaling of the original polynomial, $f(T) \rightarrow f(\lambda T)$. This scaling reflects a change in the Kondo energy, with $\lambda = T_K/T_{K0}$, where $T_{K0}$ is the zero-pressure Kondo energy. We fit the background to all other pressures by varying only the scaling factor $\lambda$. We use data between 4 and 8 K to obtain the fits, as this data is far enough from $T_C$ to not be affected by superconducting fluctuations. These fits are shown in \autoref{fig:bgsub}d, e and f. \autoref{fig:bgsub}c shows $T_K/T_{K0}$ obtained from fitting as function of pressure---we see that it smoothly decreases as a function of increasing pressure, as is expected from the qualitative behavior of the high temperature data in \autoref{fig:bgsub}a and is consistent with other studies of the Kondo energy scale in \ute \cite{kimTuningMagneticEnergy2025}. Note that at 0.77 GPa, the Kondo temperature scale is low enough that it starts to change the structure of the data at higher temperatures---this causes the background to deviate above $\approx 8$ K. 

\section{Sample 2 data}

\begin{figure}[h]
	\centering
	\includegraphics[width=0.4\textwidth]{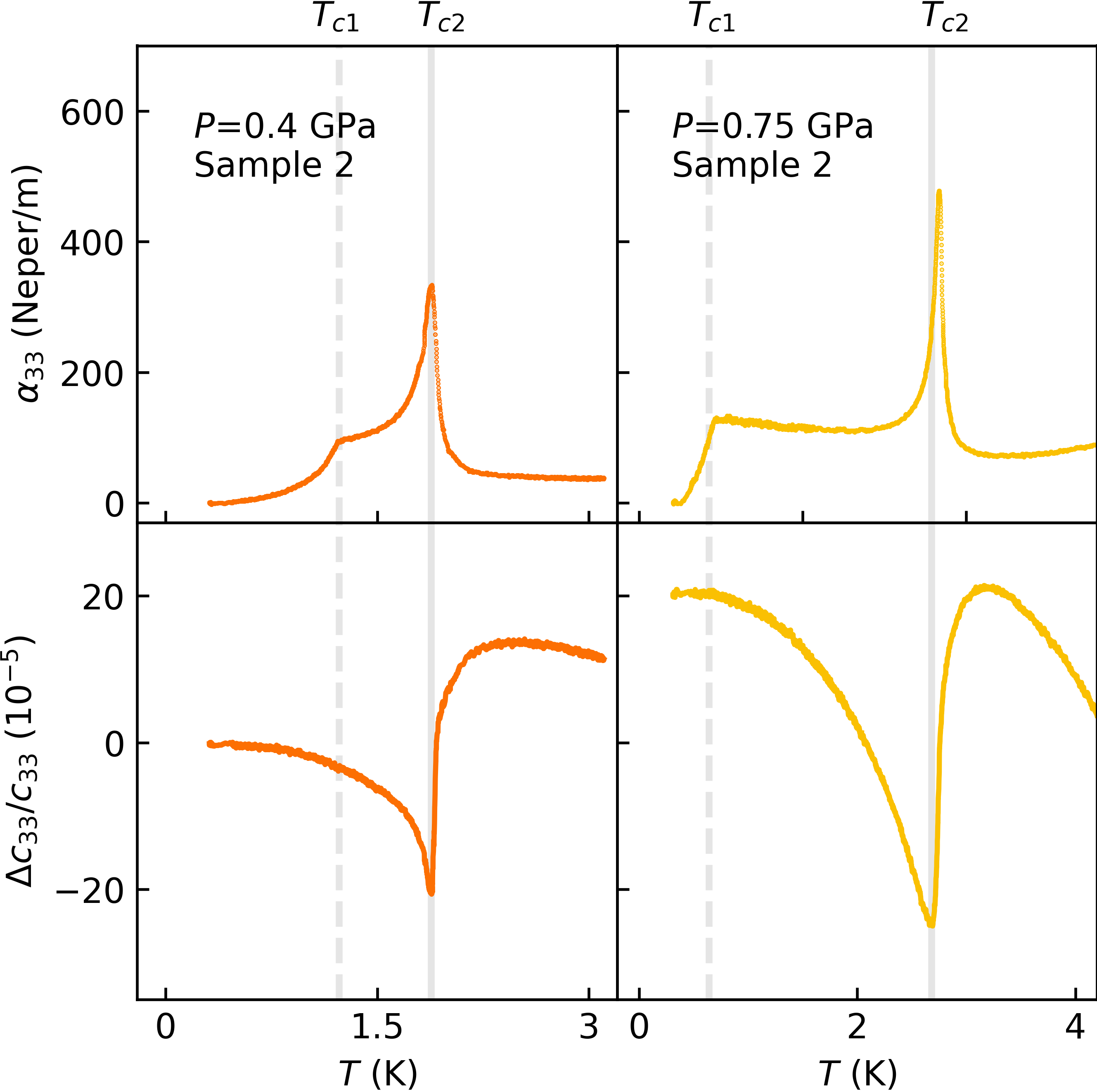}
	\caption{Elastic modulus (\ctt) and attenuation (\att) data at two different pressures on sample 2. We observe features similar to those observed in sample 1 in the main text.}
	\label{fig:sample2data}
\end{figure}

To confirm the reproducibility of the physics presented in the main text, we measure a second sample in the pressure range at which we see large precursor softening in \ctt (\autoref{fig:sample2data}). The data is similar to that measured in sample 1, with all salient features still present: 
\begin{itemize}
	\item A peak in \att at \Tct that increases with pressure, with a larger attenuation in the SC2 state than the normal state at 0.75 GPa that increases with decreasing temperature.
	\item A drop in attenuation at \Tco.
	\item Large precursor softening in \ctt with a diverging contribution that increases with pressure.
\end{itemize}

\section{Elastic Moduli in the Presence of Gaussian Fluctuations}

\subsection{Saddle point solution}

For a system close to a phase transition, we write a Landau Ginzburg Hamiltonian as a function of the (spatially non-uniform) order parameter $\vec{\psi}(\vec{x})$. For brevity, we drop the vector superscript in the following analysis, but our analysis is generally applicable to multi-dimensional order parameters.
\begin{equation}
	\mathcal{H}[\psi\left(x\right)] = \int d^3 x \left[ \alpha \, |\psi(x)|^2 + \frac{\beta}{2}\, |\psi(x)|^4 + \frac{1}{2} K \big| \nabla \psi(x) \big|^2 - \gamma \epsilon_{zz}|\psi(x)|^2\right].
\end{equation}
We define $\alpha = \alpha_0(T - T_c)$, and $\beta$ is a constant to first order in temperature. $\gamma \equiv \alpha_0 \frac{d \Tc}{d \epsilon_{zz}}$ is the strain-superconductivity coupling constant, and $K$ is a stiffness constant for variations in the order parameter. Note that $K$ can be anisotropic in principle, requiring a different stiffness for each vector component of $\nabla\psi$. Since this does not change the physics of the problem, however, we assume an isotropic $K$. Defining $\alpha' = \alpha - \gamma\epsilon_{zz}$, we can combine the terms quadratic in $\psi(x)$ to write the Hamiltonian as
\begin{equation}
	\mathcal{H}[\psi\left(x\right)] = \int d^3 x \left[ \alpha' \, |\psi(x)|^2 + \frac{\beta}{2}\, |\psi(x)|^4 + \frac{1}{2} K \big| \nabla \psi(x) \big|^2 \right].
	\label{eq:Hamiltonian}
\end{equation}

The partition function for such a Hamiltonian is:
\begin{equation}
	Z =\text{tr} \left[ e^{\mathcal{H}/\kbT} \right] = \int \mathcal{D}\psi(x) \exp{( \mathcal{H}[\psi(x)]/\kbT)}.
	\label{eq:partition_func}
\end{equation}
Here $\int \mathcal{D}\psi(x)$ denotes an integral over all the possible configurations of the order parameter $\psi(x)$

We first solve this in the mean-field or saddle-point approximation, i.e. assuming that $\psi(x)$ equals a constant $\bar{\psi} $:
\begin{equation}
	\mathcal{H}[\bar{\psi}] = V \left( \alpha' |\bar{\psi}|^2 + \frac{\beta}{2} |\bar{\psi}|^4 \right).
\end{equation}

The free energy for such a Hamiltonian is given by:
\begin{equation}
	F = -\kbT \ln Z = -\kbT \ln \left( e^{-\mathcal{H}[\bar{\psi}]/\kbT} \right) =  V \left( \alpha' |\bar{\psi}|^2 +  \frac{\beta}{2} |\bar{\psi}|^4 \right)\big|_\text{min}.
\end{equation}
This has the minima 
\[
|\bar{\psi}|^2 = 
\begin{cases} 
	0 & \text{for } \alpha'>0\\
	-\frac{\alpha'}{\beta} & \text{for } \alpha'<0.
\end{cases}
\]
This gives us for the free energy density $f = F/V$:
\begin{equation}
	f = \begin{cases}
		0 &\text{for } \alpha' >0  \\
		-\frac{\alpha'^2}{2 \beta} &\text{for } \alpha'<0
	\end{cases}
	\label{eq:SPfreeng}
\end{equation}
We can use this to find the specific heat in the mean-field approximation:
\begin{equation}
	C_{\text{MF}} = -T \frac{\partial^2 f}{\partial T^2} = \begin{cases}
		0 & \alpha' > 0\\
		-T\frac{\partial^2}{\partial T^2} \left( -\frac{ \alpha_0^2 \left(T - \Tc\right)^2}{2 \beta} \right) & \alpha'<0
	\end{cases}
\end{equation}

\begin{equation}
	\Delta C_{\text{MF}} = \frac{ \alpha_0^2 T_c}{\beta}.
	\label{eq:SPheatcap}
\end{equation}
Here $\Delta C_{\text{MF}}$ is the mean-field specific heat jump at \Tc. The elastic constant jump is obtained by a similar calculation:
\begin{equation}
	c_{\text{33,MF}} =  \frac{\partial^2 f}{\partial \epsilon_{zz}^2} = \begin{cases}
		0 & \alpha' > 0\\
		\frac{\partial^2}{\partial \epsilon_{zz}^2} \left( -\frac{ \gamma^2\epsilon_{zz}^2}{2 \beta} \right) & \alpha' < 0
	\end{cases}
\end{equation}
\begin{equation}
	\Delta c_{\text{33,MF}} = -\frac{ \gamma^2}{\beta}.
	\label{eq:SPelasticjump}
\end{equation}
Since $\gamma = \alpha_0 \frac{d\Tc}{d\epsilon_{zz}}$, we can combine \autoref{eq:SPheatcap} and \autoref{eq:SPelasticjump} to obtain the familiar Ehrenfest's relation:
\begin{equation}
	\Delta c_{\text{33,SP}} = -\left(\frac{d\Tc}{d\epsilon_{zz}}\right)^2 \frac{\Delta C_{\text{SP}}}{\Tc}.
	\label{eq:ehrenfest}
\end{equation}

\subsection{Fluctuation terms}

We now allow for small spatial variations of the order parameter, perturbing around the mean-field solution for $\bar{\psi}$:
\begin{equation}
	\psi(x) = \bar{\psi} + \phi_t (x) \hat{\epsilon_t} + \phi_l (x) \hat{\epsilon_l}.
\end{equation}
Here $\phi_t (x) \hat{\epsilon_t}$ represents fluctuations transverse to the order parameter direction and $\phi_l (x)\hat{\epsilon_l}$ longitudinal to it. In principle, $\hat{\epsilon_t}$ can be a multicomponent vector.

We assume the fluctuations are small, and hence perturb around the saddle point to second order. When they become large, our theory is not self consistent anymore, and \autoref{eq:partition_func} has to be solved exactly to obtain a free energy. To second order in small fluctuations, we get the Hamiltonian 

\begin{equation}
	\mathcal{H} = V\left( \alpha' \bar{\psi}^2 + \frac{\beta}{2} \bar{\psi}^4 \right) + 
	\int \frac{d^3 x}{2} \left( K( \nabla \phi_t)^2 + ( 2\alpha' + 2\beta\bar{\psi}^2)\phi_t^2\right) +
	\int \frac{d^3 x}{2} \left( K( \nabla \phi_l)^2 + ( 2\alpha' + 6\beta\bar{\psi}^2)\phi_l^2\right).
	\label{eq:perturbation_hamiltonian}
\end{equation}

\subsubsection{Fluctuation corrections for $T > T_C$}

The saddle point order parameter $\bar{\psi}$ is zero for $\alpha'>0$, and so $ \alpha' + 3\beta\bar{\psi}^2=\alpha' + \beta\bar{\psi}^2 = \alpha'$. The partition function for the perturbed Hamiltonian becomes:
\begin{equation}
	\begin{aligned}
		Z = \int\mathcal{D}\phi_t \exp\left(- \int \frac{d^3 x}{2} \left( K( \nabla \phi_t)^2 + 2\alpha' \phi_t^2\right)/\kbT\right) \times
		\int\mathcal{D}\phi_l \exp\left(- \int \frac{d^3 x}{2} \left( K( \nabla \phi_l)^2 + 2\alpha' \phi_l^2\right) /\kbT\right). 
	\end{aligned}
	\label{eq:partition}
\end{equation}

The kernels of the Gaussian integrals are diagonalized by the Fourier transformation
\begin{equation}
	\phi(x) = \int_q d^3q  \frac{e^{iqx}}{\sqrt{V}}\tilde{\phi}(q).
\end{equation}
The partition function in the Fourier basis becomes
\begin{equation}
	\begin{aligned}
		Z =
		&\int\mathcal{D}\tilde{\phi_t} \exp\left(- \int \frac{d^3 q}{2(2\pi)^3} \ \left(\frac{K q^2 + 2\alpha'}{\kbT} \right)\tilde{\phi_t}^2\right) \times
		\int\mathcal{D}\tilde{\phi_l} \exp\left(- \int \frac{d^3 q}{2(2\pi)^3} \ \left(\frac{K q^2 + 2\alpha'}{\kbT}\right)\tilde{\phi_l}^2\right). 
	\end{aligned}
\end{equation}
This is a standard Gaussian integral that can then be solved to obtain the free energy density $f$ \cite{kardarStatisticalPhysicsFields2007a}: 
\begin{equation}
	\begin{aligned}
		f = -\kbT \frac{\ln(Z)}{V} &= \kbT \left( \frac{1}{2} \int \frac{d^3q}{(2 \pi)^3} \ln\left(\frac{K q^2 + 2\alpha'}{\kbT}\right) 
		+ \frac{n-1}{2} \int \frac{d^3q}{(2 \pi)^3} \ln\left(\frac{K q^2 + 2\alpha'}{\kbT}\right). \right)
	\end{aligned}
\end{equation}

We assume an $n$ component order parameter, leading to $n-1$ components for the transverse fluctuations $\phi_t$.
The integrands are equal in the $\alpha'>0$ case, giving us:
\begin{equation}
	f = \frac{n \kbT}{2} \int \frac{d^3q}{(2 \pi)^3} \ln\left(\frac{K q^2 + 2\alpha'}{\kbT}\right). 
\end{equation}
Since the mean-field order parameter is zero for $T>\Tc$, this term corresponds the corrections to the free energy coming from spatial fluctuations of the order parameter. We can now calculate the fluctuation corrections to the specific heat as we approach $T_c$ from above, using $\alpha' = \alpha_0(T-T_c) - \gamma\epsilon_{zz}$:
\begin{equation}
	\begin{aligned}
		C_F\big|_{T>\Tc} = -T \frac{\partial^2 f}{\partial T^2} 
		= 2n\kb T^2 \alpha_0^2 \times \frac{1}{8\pi\big|K^{3/2}\big|\sqrt{2\alpha_0(T-T_c)}}. \\
	\end{aligned}
\end{equation}

A similar calculation is performed for the fluctuation corrections to the elastic constant:
\begin{equation}
	\begin{aligned}
		c_{33,F}\big|_{T>\Tc} = \frac{\partial^2 f}{\partial \epsilon_{zz}^2} 
		= -2n\kb T \gamma^2 \times \frac{1}{8\pi\big|K^{3/2}\big|\sqrt{2\alpha_0(T-T_c)}}
	\end{aligned}
\end{equation}

\subsubsection{Fluctuation corrections for $T < T_C$}

For $T <\Tc$, the equilibrium value of the order parameter $|\bar{\psi}|^2 = -\alpha'/\beta$. In this case, the perturbed Hamiltonian of \autoref{eq:perturbation_hamiltonian} becomes
\begin{equation}
	\mathcal{H} = V\left( \alpha' \bar{\psi}^2 + \frac{\beta}{2} \bar{\psi}^4 \right) + 
	\int \frac{d^3 x}{2} \left( K( \nabla \phi_t)^2\right) +
	\int \frac{d^3 x}{2} \left( K( \nabla \phi_l)^2 - 4\alpha'\phi_l^2\right).
\end{equation}
Note that the energetics of longitudinal (magnitude) and transverse (phase) fluctuations are now different. In particular, there are no temperature dependent terms in the integral for transverse fluctuations, and hence they do not contribute to the specific heat. We can perform Gaussian integrals similar to those in the previous section to obtain the free energy density:
\begin{equation}
	\begin{aligned}
		f = -\frac{\kbT \ln(Z)}{V} &= \alpha' \bar{\psi}^2 + \frac{\beta}{2} \bar{\psi}^4 \\
		&+ \kbT \left(\frac{1}{2} \int \frac{d^3q}{(2 \pi)^3} \ln\left(Kq^2 - 4\alpha'\right) 
		+ \frac{n-1}{2} \int \frac{d^3q}{(2 \pi)^3} \ln\left(Kq^2\right) \right).
	\end{aligned}
\end{equation}
The fluctuation part of the specific heat becomes
\begin{equation}
	\begin{aligned}
		C_F\big|_{T<\Tc} = -T \frac{\partial^2 f}{\partial T^2} 
		&= {8 k_bT_c^2 \alpha_0^2} \times \frac{1}{8\pi\big|K^{3/2}\big|\sqrt{4\alpha_0(T_c-T)}}.
	\end{aligned}
\end{equation}
The fluctuation part of the elastic modulus becomes
\begin{equation}
	\begin{aligned}
		c_{33,F}\big|_{T<\Tc} = \frac{\partial^2 f}{\partial \epsilon_{zz}^2} 
		&=  -{8 k_bT_c \gamma^2} \times \frac{1}{8\pi\big|K^{3/2}\big|\sqrt{4\alpha_0(T_c-T)}}.
	\end{aligned}
\end{equation}

We can now write both the saddle point and fluctuation contributions to the specific heat jump:
\begin{equation}
	\begin{aligned}
		C\big|_{T>\Tc} &= C_F\big|_{T>\Tc} =  \frac{1}{\sqrt{T -T_c}} \frac{n \kb T_c^2 \alpha_0^{3/2}}{4 \sqrt{2} \pi \big| K^{3/2} \big|} \\ 
		C\big|_{T<\Tc} &=C_{MF} + C_F\big|_{T<\Tc} = \frac{\alpha_0^2 \Tc}{\beta}+ \frac{1}{\sqrt{T_c - T}} \frac{\kb T_c^2 \alpha_0^{3/2}}{2 \pi \big| K^{3/2} \big|}; \\ 
	\end{aligned}
	\label{eq:heatcapfull}
\end{equation}
and the elastic constant jump:
\begin{equation}
	\begin{aligned}
		c_{33}\big|_{T>\Tc} &=c_{33,F}\big|_{T>\Tc} =  \frac{-1}{\sqrt{T -T_c}} \frac{n \kb T_c\gamma^2\alpha_0^{-1/2}}{4 \sqrt{2} \pi \big| K^{3/2} \big|} \\ 
		c_{33}\big|_{T<\Tc} &= c_{33,MF} + c_{33,F}\big|_{T<\Tc} =  \frac{-\gamma^2}{\beta}+\frac{-1}{\sqrt{T_c - T}} \frac{\kb T_c \gamma^2\alpha_0^{-1/2}}{2 \pi \big| K^{3/2} \big|}. \\ 
	\end{aligned}
	\label{eq:elasticfull}
\end{equation}
A closer inspection of \autoref{eq:heatcapfull} and \autoref{eq:elasticfull} shows that the Ehrenfest relation from \autoref{eq:ehrenfest} still holds even for non mean-field physics. The two are just related by a multiplicative constant, and thus although the following section will discuss softening in the elastic modulus, the same fluctuations should be observable in a specific heat measurement. 

\subsection{Ginzburg number}

\begin{figure}[h]
	\centering
	\includegraphics[width=0.6\textwidth]{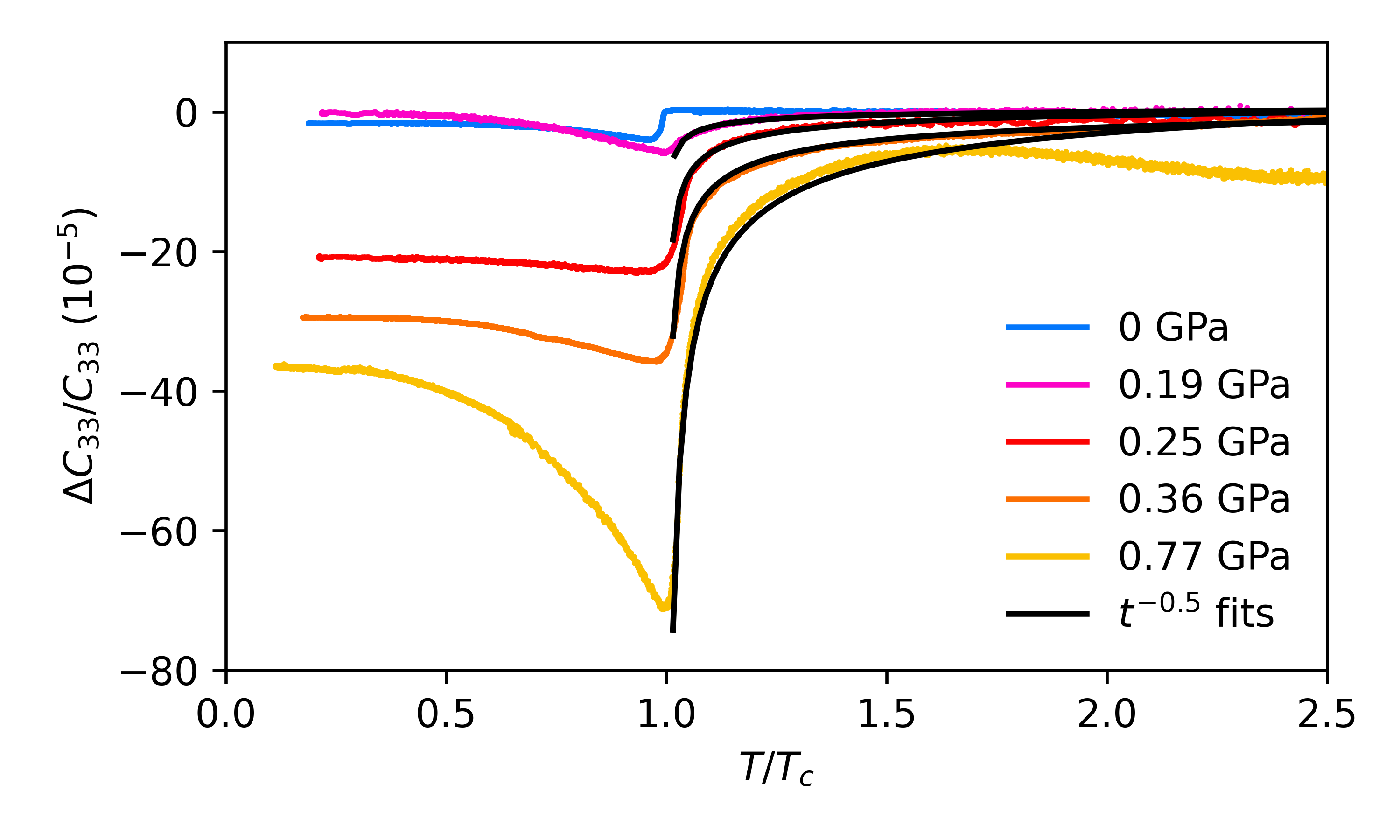}
	\caption{Background subtracted \ctt data as a function of $T/\Tc$. The black lines are fits to $a t^{-0.5}$, the form the fluctuation contribution assumes in the limit of small fluctuations. The prefactor $a$, combined with the mean-field \ctt jump is used to extract a Ginzburg number. }
	\label{fig:Gifits}
\end{figure}

We can re-write \autoref{eq:elasticfull} in terms of physical observables instead of Landau-Ginzburg parameters: 
\begin{equation}
	\Delta c_{33} = 
	-\left( \frac{d T_c}{d \epsilon_{zz}} \right)^2 \frac{\Delta C_{\text{MF}}}{T_C} \times 
	\begin{cases}
		\hspace{2.4em} \frac{n}{2\sqrt{2t}} \sqrt{\Delta C_{\text{MF}}\Tc\left(\frac{\kb\Tc}{2\pi}\right)^2} & T>\Tc \\
		\left(1 + \frac{1}{\sqrt{t}}\sqrt{\Delta C_{\text{MF}}\Tc \left(\frac{\kb\Tc}{2\pi}\right)^2} \right) & T<\Tc.
	\end{cases}
	\label{eq:mfjump2_supp}
\end{equation}
The jump in elastic modulus can be separated into two parts. First, a mean field jump turns on abruptly when $T<T_c$ at the onset of long range order and is given by $-\left( \frac{d T_c}{d \epsilon_{zz}} \right)^2 \left(\frac{\Delta C_{\text{MF}}}{T_C}\right)$. Second, a contribution that diverges as $1/\sqrt{t}$ turns on for both $T<T_c$ and $T>T_c$. We \textit{define} the Ginzburg number $Gi$ to be the reduced temperature at which the fluctuation contribution in the superconducting state equals the mean-field jump:
\begin{equation}
	Gi = \frac{ \Delta C_{\mathrm{MF}} \Tc}{\kappa_0^3} \left(\frac{k_B \Tc}{2\pi}\right)^2.
	\label{eq:ginzburg_supp}
\end{equation}
This definition is consistent with previous treatments of fluctuation phenomena in superconductors \cite{larkinTheoryFluctuationsSuperconductors2005,blatterVorticesHightemperatureSuperconductors1994,brandtFluxlineLatticeSuperconductors1995}, and it allows us to interpret the Ginzburg number as the width of the fluctuation region in reduced temperature. Written in terms of $\Delta c_{33,MF}$ and $Gi$, \autoref{eq:mfjump2_supp} becomes 
\begin{equation}
	\Delta c_{33} = 
	\Delta c_{33,MF} \times 
	\begin{cases}
		\hspace{2.4em} \sqrt{\frac{Gi}{2t}} & T>\Tc \\
		\left(1 + \sqrt{\frac{Gi}{t}} \right) & T<\Tc.
	\end{cases}
	\label{eq:mfjump3}
\end{equation}
We extract Ginzburg numbers by fitting this model to the data in \autoref{fig:Gifits}. There are just two free parameters since it is only the fluctuation terms that contribute to the elastic modulus change in the normal state: $Gi$ is obtained by fitting the data for $T>\Tc$ to $\sqrt{Gi/2t}$ (\autoref{eq:mfjump2}), and $\Delta c_{33,\text{MF}}$ is obtained from the difference in $\Delta c_{33}$ at $T/\Tc = 0$ and $T/\Tc = 2$. 

\end{document}